\documentclass[12pt]{article}

\usepackage[latin1]{inputenc}
\usepackage{parskip}
\usepackage{graphicx}
\usepackage{amstext}
\usepackage{amsmath}
\usepackage{amsfonts}
\usepackage{dcolumn}
\usepackage{dsfont}
\usepackage{epsfig}
\usepackage{longtable,lscape}
\usepackage{rotating}
\usepackage{calc}
\usepackage{ifthen}
\usepackage{threeparttable}
\usepackage{pifont}
\usepackage{geometry}
\usepackage{float}
\usepackage{txfonts}

\usepackage[none]{hyphenat}

\usepackage{caption}

\usepackage[latin1]{inputenc}
\usepackage{parskip}
\usepackage{graphicx}
\usepackage{amstext}
\usepackage{amsmath}
\usepackage{amsfonts}
\usepackage{dcolumn}
\usepackage{dsfont}
\usepackage{epsfig}
\usepackage{longtable,lscape}
\usepackage{rotating}
\usepackage{calc}
\usepackage{ifthen}
\usepackage{threeparttable}
\usepackage{natbib}
\usepackage{pifont}
\usepackage{geometry}
\usepackage{txfonts}

\usepackage{setspace}

\usepackage{algorithm}  
\usepackage{algpseudocode}
\usepackage[latin1]{inputenc}
\usepackage{graphics}
\usepackage{graphicx}
\usepackage[english]{babel}
\usepackage{tabularx}
\usepackage{lscape}
\usepackage{multirow}
\usepackage{subfig}
\usepackage{bbm}
\usepackage{calc}
\usepackage{ifthen}
\usepackage{booktabs}   

\begin{document}
	
	\bibliographystyle{elsarticle-harv}
	
\title{Nearest Neighbor Imputation for Categorical Data by  Weighting of Attributes}
	\date{}  
	\maketitle

	\vspace{-20mm}\textbf{Shahla Faisal$^{1*}$ and Gerhard Tutz$^2$}\\
	\textit{$^{1*}$ Department of Statistics, Ludwig-Maximilians-Universit\"{a}t M\"{u}nchen, Ludwigstrasse 33, D-80539, Germany.\\
		$^2$ Department of Statistics, Ludwig-Maximilians-Universit\"{a}t M\"{u}nchen, Akademiestrasse 1, D-80799 Munich, Germany.\\
		$^*$Correspondence E-mail: shahla.ramzan@stat.uni-muenchen.de}\vspace{10mm}\\
	{\textbf{Summary}}\\
	\hspace*{25pt} 
	{	Missing values are a common phenomenon in all areas of applied research. While various imputation methods are available for metrically scaled variables, methods for categorical data are scarce. An imputation method that has been shown to work well for high dimensional metrically scaled variables is the imputation by nearest neighbor methods. In this paper, we extend the  weighted nearest neighbors approach to impute missing values in categorical variables. The proposed method, called $\mathtt{wNNSel_{cat}}$, explicitly uses the information on association among attributes. 
	The performance of different imputation  methods is compared in terms of the proportion of falsely imputed values. Simulation results show that the weighting of attributes yields smaller imputation errors than existing approaches. A variety of real data sets is used to support the results obtained by simulations.} \vspace{5mm}\\
	\emph{Keywords:} Attribute weighting; Categorical data; Weighted nearest neighbors; Kernel function; Association.

\onehalfspacing

\section{Introduction}
Categorical data are important in many fields of research, examples are surveys with multiple-choice questions in the social sciences \citep{chen2000nearest}, single nucleotide polymorphisms (SNPs) in genetic association studies \citep{schwender2012imputing} and tumor or cancer studies \citep{eisemann2011imputation}. It is most likely that some respondents/patients do not provide the complete information on the queries, which is the most common reason for missing values.
Sometimes, also the information may not be recorded or included into the database. Whatever the reason, missing data  occur in all areas of applied research. Since for many statistical analyses a complete data set is required, the imputation of missing values is a useful tool.
For categorical data, although prone to contain missing values, imputation tools  are scarce.

It is well known that using the information from complete cases or available cases may lead to invalid statistical inference \citep{little2014statistical}. A common approach is to use  an appropriate imputation model, which accounts for the scale level of the measurements.  When the data are categorical the log-linear model is an  appropriate choice \citep{schafer1997analysis}. The simulation studies  of \cite{ezzati1995simulation} and \cite{schafer1997analysis} showed that it provides an attractive solution for missing categorical data problems.
However, its use is restricted to cases with a small number of attributes  \citep{erosheva2002alternative} since model selection and fitting becomes very challenging for larger dimensions.

A non-parametric method called hot-deck imputation has been proposed as an alternative \citep{rubin1987multiple}.
This technique searches for the complete cases having the same values on the observed variables as the case with missing values. The imputed values are drawn from the empirical distribution defined by the former. The method is well suited even for  data sets with a large number of attributes \citep{cranmer2013we}. A variant, called approximate Baysian bootstrap, works well in situations where the standard hot-deck fails to provide proper imputation \citep{RubinSchenker1986}.
But the hot-deck imputation may yield biased results irrespective of the missing data mechanism \citep{SchaferGramham2002}, and it may become less likely to find  matches if the number of variables is large \citep{andridge2010review}.

Another popular non-parametric approach to impute missing values is the nearest neighbors method \citep{troyanskaya:2001}. The relationship among attributes is taken into account when computing the degree of nearness or distance. The method may easily be implemented for high-dimensional data. However, the $\mathsf{k}$-nearest neighbors ($\mathsf{k}$NN) method, originally developed for continuous data, cannot be employed without modifications to non-metric data such as nominal or ordinal categorical data \citep{schwender2012imputing}. As the accuracy of the $\mathsf{k}$NN method is mainly determined by the distance measure used to calculate the degree of nearness of the observations, 
one needs different distance formula when data are categorical.
Some existing methods to impute attributes are based on the mode or weighted mode of nearest neighbors \citep{liao2014missing}.

\cite{schwender2012imputing}  suggested a weighted $\mathsf{k}$NN method to impute \textit{categorical} variables only, that uses the Cohen or Manhattan distance for finding the nearest neighbors. The imputed value is calculated by using weights that correspond  to the inverse of the distance. One limitation of this approach is that it can handle only variables that have the same number of categories. Also the value of $\mathsf{k}$, which strongly affects the imputation estimates, is  needed. There are some methods for imputing mixed data that can also be used for categorical data, for example, see \cite{liao2014missing} and \cite{stekhoven2012missforest}. The latter transform  the categorical data to dichotomous data and use the classical $\mathsf{k}$-nearest neighbors method to the standardized data with mean 0 and variance 1. The imputed data are re-transformed to obtain the estimates. 
However, it has been confirmed by several studies that rounding may lead to serious bias, particularly in regression analysis (\citet{allison2005imputation}, \citet{horton2003potential}).

For categorical data  one has to use specific  distances or similarity measures, which are typically based on contingency tables.
Commonly used distance measures include the simple matching coefficient, Cohen's kappa $\kappa_c$ \citep{cohen1960kappa}, and the Manhattan or $L_1$ distance.
The Euclidean or variants of the Minkowski distance give an equal importance to all the variables in the data matrix when computing the distance. But for a larger number of variables, the equal weighting ignores the complex structure of correlation/association among these variables.
As will be demonstrated, better distance measures are obtained by utilizing the association between variables.
More specific, we propose a weighted distance that explicitly takes the association among covariates into account. Strongly associated covariates are given higher weights forcing them to contribute more strongly to the computation of the distances than weakly associated covariates.

The paper is organized as follows: Section \ref{sec:distcategorical} reviews some available measures to calculate the distance for nominal data. The improved measure, which uses information on association among attributes, is introduced. A weighted nearest neighbors procedure is  described in Section \ref{sec:wNNcatSelect}.
In Section \ref{EvalExist}, the existing methods to impute missing categorical data, and the measure of performance used for comparison are described. In section \ref{sec:simul} the performance of different imputation methods is compared in simulation studies. Section \ref{sec:appli} gives applications of the proposed method to some real data sets. 

\section{Methods}
\label{sec:distcategorical}

At the core of nearest neighbor methods is the definition of the distance.
In contrast to continuous data, computation of distance or similarity between categorical data is not straightforward since categorical values have no explicit notion of ordering. 
We first consider distances for categorical variables that can be used to impute missing values.

\subsection{Distances for Categorical Variables}

Let data be collected in a $(n \times p)$-matrix $\mathbf{Z} =(Z_{is})$, where $Z_{is}$ is the $i^{th}$ observation on the $s^{th}$ attribute.
Let  $\mathbf{z}_i^T = (Z_{i1}, \cdots, Z_{ip})$ denote the $ith$ row or observation vector in the data matrix $\mathbf{Z}$. 
The categorical observations  $Z_{is}$ in the data matrix $\mathbf{Z}$  can take values from  $ \{1,\dots,k_s\}, ~s=1,\dots,p$, where $k_s$ is the number of categories of the $s^{th}$ attribute. Distances can use the original variables $Z_{is} \in {1,\dots, k_s}$ or the vector representation.
It is to note here that '$\mathsf{k}$' denotes the number of nearest neighbors, whereas '$k$' (or '$k_s$') corresponds to the number of categories that an attribute in the data matrix $\mathbf{Z}$ may assume.

\subsubsection*{Simple Matching of Coefficients (SMC)}
The distance uses the original values $Z_{is}$. 
This method   simply considers the matching of  the values of the variables, that is, whether they are the same or not \citep{sokal1958statistical}. The SMC distance between two observation vectors $\mathbf{z}_i,\mathbf{z}_j$ is defined by
\[
d_{SMC}(\mathbf{z}_i,\mathbf{z}_{j}) = \sum_{s=1}^{p} I({ Z_{is} \neq Z_{js}})
\]

where $I(.)$ is an indicator function defined by

\begin{equation*}
I(.) = \left\{\begin{array}{ll}
1    &\       \text{if } Z_{is} \neq Z_{js}  \\
0   &\ \text{otherwise} .
\end{array}\right.
\end{equation*}

\subsubsection*{Minkowski's Distance}

The Minkowski's distance can be used for re-coded categorical variables.

For the computation  the categorical variable $Z_{is}$ is transformed into binary variables.
Let ${\mathbf{z_{is}}^T}=(z_{is1},\dots,z_{isk_s})$ be the dummy vector built from $Z_{is}$ with components being defined by
\begin{equation*}
z_{isc} = \left\{\begin{array}{ll}
1    &\       \text{if } Z_{is} = c, \\ 
0   &\ \text{otherwise} .
\end{array}\right.
\end{equation*}
Let $\mathbf{Z}^D$ denote the matrix of dummies which is obtained from the original data matrix. Thus, the $ith$ row of the matrix $\mathbf{Z^D}$ has the form   $(\mathbf{z}_{i1}^T, \cdots,\mathbf{z}_{ip}^T)^T$ with dummy vectors $\mathbf{z}_{is}$, $s=1,\cdots,p$.

The dummy vectors ${\mathbf{z_{is}}^T}$ for a nominal variable with four categories can be written as    
\begin{center}
	$$ \begin{array}{ccccc}
	$category$ & $$z_{is1}$$ & $$z_{is2}$$& $$z_{is3}$$& $$z_{is4}$$\\
	1& 1& 0& 0& 0\\
	2& 0& 1& 0& 0\\
	3& 0& 0& 1& 0\\
	4& 0& 0& 0& 1\\
	\end{array}
	$$
\end{center}

By using the dummy vectors from $\mathbf{Z^D}$ the Minkowski's distance between two rows $\mathbf{z}_i,\mathbf{z}_j$ of $\mathbf{Z}$ is given by

\begin{equation}
d_{Cat}(\mathbf{z}_i,\mathbf{z}_j)= \left(  \sum_{l=1}^{p}  \sum_{c=1}^{k_l} |z_{ilc}-z_{jlc}|^q  \right) ^{1/q},
\label{eq:catDistq}
\end{equation}

By choosing $q=2$, one obtains the  Euclidean distance

\begin{equation*}
d_{Cat}(\mathbf{z}_i,\mathbf{z}_j)= \left(  \sum_{l=1}^{p}  \sum_{c=1}^{k_l} (z_{ilc}-z_{jlc})^2  \right) ^{1/2},
\label{eq:catDist2}
\end{equation*}

and for $q=1$ one obtains  the Manhattan or $L_1$ distance

\begin{equation*}
d_{Cat}(\mathbf{z}_i,\mathbf{z}_j)=   \sum_{l=1}^{p}  \sum_{c=1}^{k_l} |z_{ilc}-z_{jlc}|.
\label{eq:catDist1}
\end{equation*}

Thus the Euclidean distance and Manhattan distance are two special forms of the Minkowski's distance.
It should be noted that it has a strong connection to the matching coefficient distance.
When $q=1$, the Minkowski's distance uses the number of matches between the two covariates, ant the distance is equal to the simple matching coefficient (SMC) distance.

\subsection{Selection of Attributes by Weighted Distances }

The Euclidean or variants of Minkowski's distance give an equal importance to all the variables in the data matrix. When the number of variables is large and they are correlated/ associated, it is useful to give unequal weights to covariates when calculating the distance.
We present a weighted distance  which explicitly takes the association among covariates into account. More specifically, highly associated covariates will contribute more to the computation of the distance than less associated covariates.

For a concise definition we distinguish between  cases that  were observed in the corresponding component  and missing values, only the former contribute to the computation of the distance.
Let us again consider the data matrix $\mathbf{Z}$    with dimension $n \times p$. Let $\mathbf{O}=(o_{is})$ denote the $n \times p$ matrix of dummies, with $o_{is}=0$ if the value is missing, and $o_{is}=1$ if the the value is available in the data matrix $\mathbf{Z}$.
Let now $Z_{is}$ be a specific missing entry in the data matrix $\mathbf{Z}$, that is, $o_{is}=0$. 
For the computation of distances we use the corresponding matrix of dummy variables $\mathbf{Z}^D$. 

We propose to use as distance between the $i$-th and the $j$-th observation
\begin{equation}
d_{CatSel}(\mathbf{z}_i,\mathbf{z}_j)= \left( \frac{1}{a_{ij}} \sum_{l=1}^{p}  \sum_{c=1}^{k_l} |z_{ilc}-z_{jlc}|^q I{(o_{il}=1)}I{(o_{jl}=1)} C(\delta_{sl}) \right) ^{1/q},
\label{eq:catDist}
\end{equation}
where $I{(.)}$ denotes the indicator function and  $a_{ij}=\sum_{l=1}^p I{(o_{il}=1)}I{(o_{jl}=1)}$ is the number of valid components in the computation of distances. The crucial part in the definition of the distance is the weight $C(\delta_{sl})$. $C(.)$ is a convex function defined on the interval $[-1,1]$ that transforms the measure of association  between attributes $s$ and $l$, denoted by $\delta_{sl}$, into weights.

It is worth noting that the distance is now specific to  the $s^{th}$ attribute, which is to be imputed.  For $C(.)$ we use the power function $C(\delta_{sl})=\lvert \delta_{sl}\rvert^\omega$.
So the attributes that  have a higher association with the $s^{th}$ attribute are contributing more to the distance   and vice versa. 
The higher the value of association, the more it contributes to the computation of the distance. Note also that only the \textit{available} pairs with $I{(o_{il}=1)}I{(o_{jl}=1)}$ are used for the computation   of the distance.
In the following we describe some of association among variables that account for the number of categories each variable can take.

\subsection{Measuring Association Among Attributes} \label{Accociationmeasures}
One important issue in the computation of distances in equation (\ref{eq:catDist}) is how to compute the association ($\delta_{sl}$) among categorical variables as the usual Pearson coefficient of correlation is not suitable for categorical covariates. In this section, we briefly describe some measures that are used to calculate the association between categorical measurements.

The measures of association for two nominal or categorical variables are typically based on the $\chi^2$-statistic which tests the independence of variables in contingency tables.

Consider the association among attribute $s$ and $l$ where attribute $s$ has $i=1,...,k_s$ categories and attribute $l$ has $j=1,...,k_l$ categories.
The two attributes $s$ and $l$ can be presented in the form of an $k_s \times k_l$ contingency table (Figure \ref{fig:ContTable}).

\begin{figure}[h]
	
\sf	\centering
	
	\begin{tabular}{cc|cccccc|c}
		&	 & \multicolumn{6}{c}{Attribute $l$} & \\
		\rule[-1ex]{0pt}{1.5ex} &   & 1 & 2 & $\cdots$  & j & $\cdots$ &  $k_l$  & Total \\
		
		\hline \rule[-1ex]{0pt}{3.5ex} 	\parbox[t]{1mm}{\multirow{6}{*}{\rotatebox[origin=c]{90}{Attribute $s$}}}	&  1 & $n_{11}$ & $n_{12}$ &  &  &  &  &  $n_{1.}$\\
		& 	2 & $n_{21}$ & $n_{22}$  &  &  &  &  &  $n_{2.}$ \\
		& 	$\cdots$  &  &  &$\ddots$  &  &  &  &  \\
		& 	$i$  &  &  &  & $n_{ij}$ &  &  &  $n_{i.}$ \\
		& 	$\cdots$&  &  &  &  &$\ddots$  &  &  \\
		& 	$k_s$ &  &  &  &  &  & $n_{k_sk_l}$ &  \\
		
		\hline \rule[-1ex]{0pt}{1.5ex} & Total & $n_{.1}$ & $n_{.2}$ &  & $n_{.j}$ &  &  & $n_{}$ \\
		\hline
	\end{tabular} 	 	
	\caption{ Contingency table with $k_s\times k_l$ cells }
	\label{fig:ContTable}
\end{figure}

In the contingency table, $n_{ij}$ is the number of observations  ($Z_s=i$, $Z_l=j$) 
and $n = n_{..}$ is the total number of values.
The $\chi^2$-statistic between attributes $s$ and $l$ is defined as

\[
\chi^2_{sl} = \sum_{i,j}^{} \frac{( n_{ij} - \frac{n_{i.}n_{.j}}{n} )^2 }{ \frac{n_{i.}n_{.j}}{n} },
\]
where $n_{i.}$ and $n_{.j}$ are the row and columns totals respectively and $n$ is the total number of observations in the contingency table.

Association measures based on the $\chi^2$-statistic are, in particular, the  $\phi$-coefficient, Pearson's contingency Coefficient and Cramer's V.

\subsubsection*{Phi coefficient ($\phi$)}
For nominal variables with only two categories, i.e. $k_s = k_l = 2$, a simple measure of association is the $\phi$-coefficient
\[
\phi_{sl} = \sqrt{  \frac{\chi^2_{sl}}{n} }.
\]

\subsubsection*{Pearson's coefficient of contingency ($\mathcal{PCC}$)}
For  $k_s = k_l = k$, that is,  the variables have the same number of categories,   Pearson's coefficient of contingency is computed  as
\[
C_{sl} = \sqrt{  \frac{\chi^2_{sl}}{\chi^2_{sl} + n} }.
\]
It can be corrected to reach a maximum value of 1 by dividing by the factor  $\sqrt{(k-1)/k}$, where $k$ is the number of rows($r=k_s$) or columns ($c=k_l$) as both are equal. Pearson's corrected coefficient (PCC) of contingency is given by

\[
\text{PCC}_{sl} = \frac{C_{sl}} {\sqrt{(k-1)/k}}.
\]

It is suitable only when the number of categories of both covariates are the same.

\subsubsection*{Cohen's kappa ($\kappa_c$)}
For  $k_s = k_l = k$,  another useful measure of association was given by \cite{cohen1960kappa},

\[
\kappa_{sl} = \frac{p_0 - p_e}{1-p_e},
\]
where
$ p_0 = n_{ii}/n = n_{jj}/n $ 
are the proportions of units with perfect agreement, which are the diagonal elements in the contingency table and
$p_e = \sum_{i=j}^{k}\frac{n_{i.}n_{.j}}{n} $ is the expected proportion of units under independence.

\subsubsection*{Cramer's V}
If the covariates have different number of categories ($k_s \neq k_l$),  Cramer's V is an attractive option \citep{cramer1946methods}. It is defined by

\[
\text{Cramer's V} = \sqrt{  \frac{\chi^2_{sl}/n }{min(k_s-1,k_l-1)}  },
\]
where $n=k_s\times k_l$ is the total number of cells in the contingency table.

In this paper, we choose Cramer's V as the measure of association ($\delta_{sl}$) to be used in  (\ref{eq:catDist}) because its value lies between 0 and 1; and it can be used for unequal number of categories of the attributes as well. The corresponding method is denoted by $\mathtt{wNNSel_{cat}}$.


\section{Using Nearest Neighbors to Impute Missing Values}
\label{sec:wNNcatSelect}

Classical nearest neighbor approaches fix the number of neighbors that are used. We prefer to use weighted nearest neighbors by using weights that are defined by kernel functions. Uniform kernels yield the classical approach, however, smooth kernels typically provide better results.

Let $Z_{is} $ 
be a missing value in the $n \times p $ matrix of observations. The $\mathsf{k}$ nearest neighbor observation vectors  are defined by

\[
\mathbf{z}_{(1)}^D,\dots,\mathbf{z}_{(\mathsf{k})}^D \quad \text{with} \quad d(\mathbf{z}_i,\mathbf{z}_{(1)}) \le\dots \le d(\mathbf{z}_i,\mathbf{z}_{(\mathsf{k})})
\]

where $\mathbf{z}_{(i)}^D$ are rows from the matrix $\mathbf{Z}^D$, and $d(\mathbf{z}_i,\mathbf{z}_{(\mathsf{k})})$ is the  computed distance using equation (\ref{eq:catDist}). It is important to mention that the row $\mathbf{z}_{(i)}^D$ is composed of values of dummy variables of the form
$ ( \mathbf{z}_{(i)1}^T , \cdots, \mathbf{z}_{(i)p}^T)^T $, where $\mathbf{z}_{(i)s}^T = (z_{is1}, \cdots,  z_{isk_s})$ are the dummy values.

For the imputation of the value $Z_{is} $ we use the weighted estimator

\begin{equation} \label{eq:relprob}
{\hat \pi}_{isc} =  \sum_{j=1}^\mathsf{k}  w(\mathbf{z}_i,\mathbf{z}_{(j)}) \mathbf{z}_{(j)sc}, 
\end{equation}

with weights given by

\begin{equation} \label{eq:weight}
w(\mathbf{z}_i,\mathbf{z}_{(j)})= \frac{ K(  \frac {d(\mathbf{z}_i,\mathbf{z}_{(j)})} {\lambda})}{ \sum_{h=1}^\mathsf{k} K( \frac{d(\mathbf{z}_i,\mathbf{z}_{(h)})}  {\lambda})},
\end{equation}

where $K(.)$ is a kernel function (triangular, Gaussian etc.)  and $\lambda$ is a tuning parameter.
Note that $\mathbf{\hat \pi}_{is}^T = (\hat \pi_{is1}, \cdots, \hat \pi_{isk_s})$ is a vector of estimated probabilities. 

If one uses all the available neighbors that is $\mathsf{k}={n}$, then $\lambda$ is the only and crucial tuning parameter.
The imputed estimate of $Z_{is}$ is the value of $c \in \{1,\dots,k_s\}$ that has the largest value.
In other words, the weighted imputation estimate of a categorical missing value $Z_{is}$ is

\begin{equation} \label{eq:zhat}
\hat Z_{is}= \text{arg max}_{c=1}^{k_s} \;  {\hat \pi}_{isc},
\end{equation}

If the maximum is not unique  one value  is selected at random.
The proposed $\mathtt{wNNSel_{cat}}$ procedure can be described as follows:
\begin{enumerate}
	\item Locate a missing value $Z_{is}$ in sample $i$ and attribute $s$ in the data matrix.
	\item Compute $d_{\text{CatSel}}$ between observation vectors $\mathbf{z}_i,\mathbf{z}_j$  using equation (\ref{eq:catDist}).
	\item Rank samples (rows) of the matrix $\mathbf{Z}^D$ based on $d_{\text{CatSel}}$.
	\item Compute the corresponding weights  based on kernel function $w(\mathbf{z}_i,\mathbf{z}_{j})$ using equation (\ref{eq:weight}).
	\item Compute the s probabilities by using equation (\ref{eq:relprob})
	\item Compute the missing value $\hat{z}_{is}$ by using equation (\ref{eq:zhat}).
	\item Seek the next missing value in $\mathbf{Z}$ and repeat steps (2-6) until all missing values in $\boldsymbol Z$ have been imputed imputed.
\end{enumerate}

In this weighted nearest neighbor ($\mathtt{wNNSel_{cat}}$) method, one has to deal two tuning parameters, $\lambda$ in (\ref{eq:weight}) and $\omega$ to be used in distance equation (\ref{eq:catDist}). These tuning parameters have to be specified

\subsection{A Pearson Correlation Strategy}
As an alternative we consider a strategy that uses the dummy variables directly. 
Starting from the matrix of dummies  $\mathbf{Z}^D$ we use the Pearson correlation coefficient between dummy variables as association measure. The weighting scheme remains the same. The imputation is again determined by
\[
{\hat \pi}_{isc} =  \sum_{j=1}^\mathsf{k}  w(\mathbf{z}_i,\mathbf{z}_{(j)}) \mathbf{z}_{(j)sc}.
\]
Although ${\hat \pi}_{isc}^T=({\hat \pi}_{is1},\dots,{\hat \pi}_{isk_s})$  might not be a vector of probabilities, simple standardization by setting
\[
{\tilde \pi}_{isc} =  {\hat \pi}_{isc}/\sum_{r=1}^{k_s}{\hat \pi}_{isr}
\]
yields a vector of probabilities that can be used to determine the mode. The method can be seen as an adaptation of the weighting method proposed by \cite{TuRam2015}. Only small modification are needed to apply the method to the dummy variables. One is that the missing of an observation refers  to a set of variables, namely all the dummies that are linked to a missing value. 
For this method the Gaussian kernel and the Euclidean distance are used throughout.
The method is denoted by $\mathtt{wNNSel_{dum}}$.

\subsection{Cross Validation}

The imputation procedure $\mathtt{wNNSel_{cat}}$ requires pre-specified values of the tuning parameters $\lambda$ and $\omega$. In this section we present a cross validation algorithm that automatically choses those values for which the imputation error is minimum.

To estimate the tuning parameters $\omega$ and $\lambda$, we set some available values $(o_{is})=1$ in the data matrix as missing $(o_{is})=0$. The advantage of this step is that these values are used  to estimate the tuning parameters. The procedure of cross validation to find optimal tuning parameters is given as algorithm 1.

\begin{algorithm}[h]
	\caption{Cross-validation for \texttt{wNNSelcat}} 	\label{cv_wNNSelect}
	\begin{algorithmic}[1]
		\Require {$\mathbf Z$ an $n \times p $ matrix, number of validation sets $t$, range of suitable values for tuning parameters $\mathcal{L}$ and $\mathcal{W}$}
		\State $\mathbf Z^{\text{cv}} \leftarrow$ initial imputation using unweighted 5-nearest neighbors
		\For{ $t$ = $1, \ldots, \mathcal{T}$ }
		\State  $\mathbf Z^{\text{cv}}_{\text{miss,t}}$ $ \leftarrow $ artificially introduce missing values to $\mathbf Z^{cv}$
		\For{ $\omega \in \mathcal{W}$ }
		\For{$\lambda \in \mathcal{L}$}
		
		\State  $\mathbf Z^{\text{cv}}_{ wNNSel_{cat},t} \leftarrow $ imputation of $\mathbf Z^{\text{cv}}_{\text{miss,t}}$ using $\mathtt{wNNSel_{cat}}$ procedure
		
		\State $\psi_{(\lambda, \omega),t}  \leftarrow$ imputation error (PFC) of $\mathtt{wNNSel_{cat}}$ procedure for $\lambda$ \& $\omega$
		
		\EndFor
		\EndFor

		\State Determine $(\lambda,\omega)_{\text{best}}  \leftarrow$ \text{argmin} $\frac{1}{T} \sum^{T}_{t=1} {\psi_{(\lambda, \omega),t}}$
		\EndFor
		\State	$\mathbf Z^{\text{imp}} \leftarrow$  $\mathtt{wNNSel_{cat}}$ imputation of $\mathbf Z$ using $(\lambda,\omega)_{\text{best}} $
	\end{algorithmic}
\end{algorithm}

\section{Evaluation of Performance} \label{EvalExist}
This section briefly describes some available methods to impute missing categorical data.
Then the performance of imputation methods is compared by  using the mean proportion of falsely imputed categories (PFC) given by
\[
\text{PFC} = \frac{1}{m} \sum_{Z_{is}:o_{is}=0}  I {(Z_{is} \neq \hat Z_{is} )},
\]
where $I(.)$ is an indicator function which takes the value 1 if $Z_{is} \neq \hat Z_{is}$ and 0 otherwise, $\omega$ is the number of missing values in the data matrix, $Z_{is}$ is the true value and $\hat Z_{is}$ is its imputed value.

\subsection{Existing Methods}
In this section, we briefly review some existing procedures for the imputation of categorical missing data.

\subsubsection*{Mode Imputation}
This is perhaps the simplest and fastest method to fill the incomplete categorical data. 
The missing values of an attribute are replaced by the attribute with maximal occurrence in that variable, that is, the mode. The approach is very simple and totally ignores the association or correlation among the attributes.

\subsubsection*{k-Nearest Neighbors Imputation}
In this method, $\mathsf{k}$ neighbors are chosen based on some distance measure and their average is used as an imputation estimate. The method requires the selection of a suitable value of $\mathsf{k}$, the number of nearest neighbors, and a distance metric. The function \texttt{kNN} in the R package \texttt{VIM} \citep{RPackage:vim} can impute categorical and mixed type of variables.

An adaptation of the $\mathsf{k}$ nearest neighbors algorithm proposed by \cite{schwender2012imputing} can impute missing genotype or categorical data. The procedure selects $\mathsf{k}$ nearest neighbors based on distance measures (Cohen, Pearson, or SMC).
The weighted average of the $\mathsf{k}$ nearest neighbors is used to estimate the missing value, where the weights are defined by the inverse of the distances. The limitation of this method is that it offers imputation only for variables having an equal number of categories.
We use the function \texttt{knncatimpute} from R package \texttt{scrime} \citep{RPackage:scrime} to apply this method.

\subsubsection*{Random Forests}
In recent years,  random forest \citep{breiman2001random} have been    used  in various areas including imputation of missing values.
The imputation of missing categorical data by random forests  is based on an iterative procedure that uses initial imputations using mode imputation and then improves the imputed data matrix on successive iterations. A random forest model is developed for each predictor with a missing value by using the rest of the predictors and the model   to estimate the missing value of that predictor. The imputed data matrix is updated and the difference between previous and new imputation is assessed at the end of each iteration. The whole process is repeated until a specific criterion is met (\citealp{stekhoven2012missforest}; \citealp{rieger2010random}; \citealp{segal2004machine}; \citealp{pantanowitz2009missing}).


The main advantages of random forests include the ability to handle high dimensional mixed-type data with non-linear and complex relationships, and robustness to outliers and noise (\cite{hill2012four},  \cite{rieger2010random}). The \texttt{missForest} package in the statistical programming language \texttt{R}  offers this approach \citep{RPackage:missForest}.

\section{Simulation studies} \label{sec:simul}
This section includes preliminary simulations to check if the suggested distance measure contributes to better imputation or not. Using  simulated data we compare our method  with three existing methods in the situations with binary or multi-categorical variables only, and mixed (binary and multi-categorical) variables.

\subsection{Binary Variables}

\begin{figure}[h]
	\centering
	\begin{tabular}{ccc}
		\includegraphics[width=0.3\linewidth]{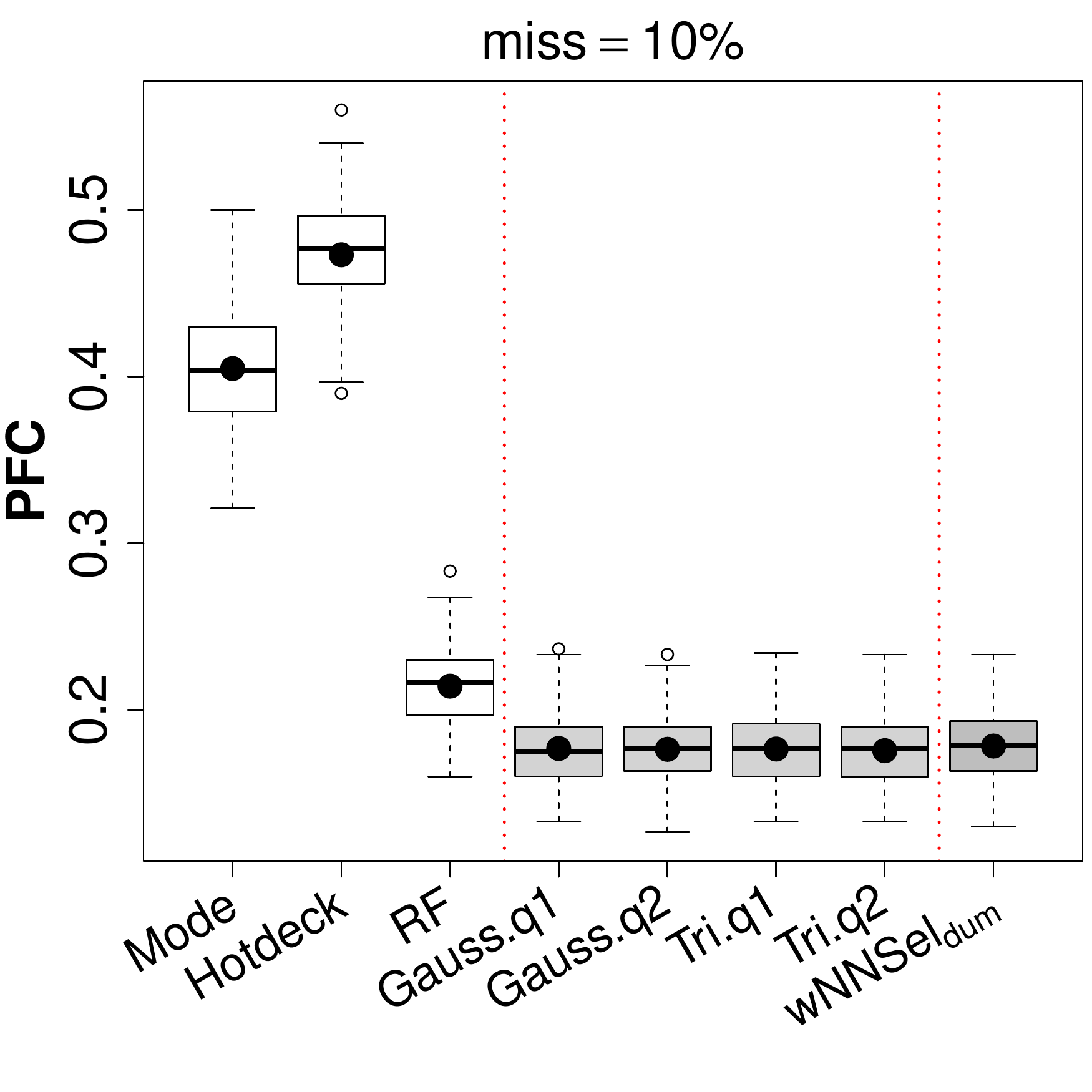} &
		\includegraphics[width=0.3\linewidth]{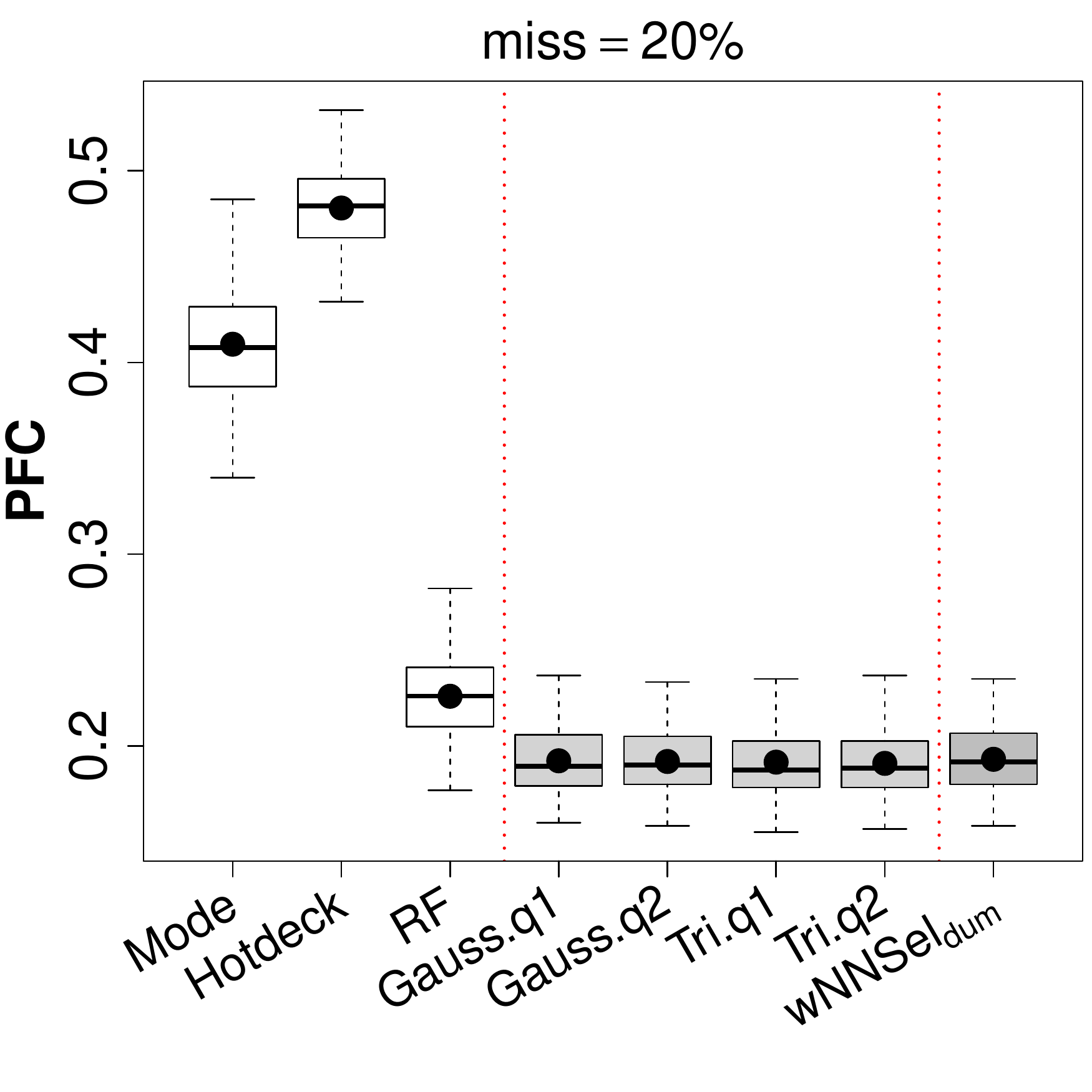} &
		\includegraphics[width=0.3\linewidth]{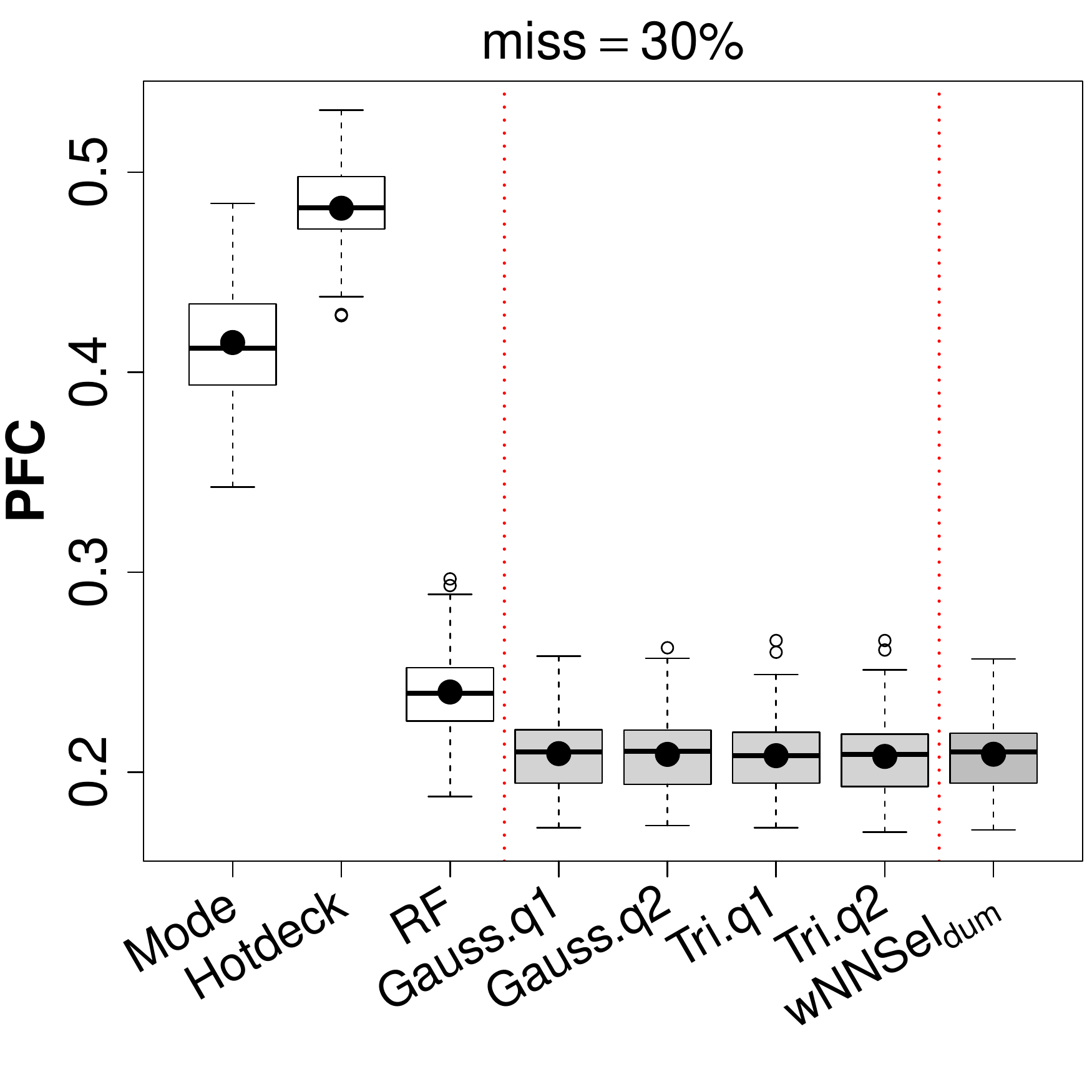}	\\
	\end{tabular}
	
	\caption{Simulation study for binary data: boxplots of PFCs for MCAR missing data with $n=100$, $p=30$. Solid circles within boxes show mean values.}
	\label{fig:fig_Bin_m10}
\end{figure}

In our fist simulation study we investigate the performance of imputation methods using simulated binary data. We generate $S=200$ samples of size $n=100$ for $p=30$ predictors drawn from a multivariate normal distribution with $N(\mathbf 0 , \boldsymbol \Sigma)$. The correlation matrix $\boldsymbol \Sigma$ has an autoregressive type of order $1$ with $\rho=0.8$. The data are converted to binary variables by defining positive values as the first category and negative as the second. In each sample randomly selected values are declared as missing with proportion of 10\%, 20\% and 30\%.
The missing values are imputed using mode imputation, random forests (\texttt{RF}) and the proposed weighted nearest imputation methods. In the weighted nearest imputation methods ($\mathtt{wNNSel_{cat}}$), the distance (\ref{eq:catDist}) is computed for $q=1,2$. We use the Gaussian and the triangular kernel functions for each value of $q$ (shown as \texttt{Gauss.q1}, \texttt{Gauss.q2}, \texttt{Tri.q1}, and \texttt{Tri.q2} in Figure \ref{fig:fig_Bin_m10}).

The resulting PFCs are shown in Figure \ref{fig:fig_Bin_m10}. It is clear from the figure that the weighted imputation methods ($\mathtt{wNNSel_{cat}}$ and $\mathtt{wNNSel_{dum}}$) yield smaller error than mode, hot deck and random forest imputation methods. The highest errors are obtained by hot deck and mode imputation. It seems that the selection of the kernel function and the value of $q$ do not affect the results. 
The average values of the PFCs are nearly equal for $q=1,2$ when using the triangular kernel. For the Gaussian kernel, $q=2$ produces slightly smaller imputation errors. Overall the ($\mathtt{wNNSel_{cat}}$ and $\mathtt{wNNSel_{dum}}$) procedures give similar results. We skip the Hotdeck method in our further simulations as it produced poor imputations.

\subsection{Multi-categorical Variables}

In this section, we investigate the performance of imputation methods using multi-categorical data.
We generate $S=200$ samples of size $n=100$ for $p=10,50$ predictors drawn from a multivariate normal distribution with $N(\mathbf 0 , \boldsymbol \Sigma)$. The correlation matrix $\boldsymbol \Sigma$ has an autoregressive type of order $1$ with $\rho=0.9$.
These values are then converted to the desired number of categories. In each sample, $miss$ = 10\%, 20\%, 30\% of the total values were replaced by missing values completely at random (MCAR).

In the case where all attributes have an equal number of categories ($k_s=k$), another benchmark proposed by \cite{schwender2012imputing}, is also considered. To compare the performance, the proportion of falsely imputed categories (PFC) is computed for each imputation method. We distinguish the cases when the probabilities are the same for all categories and the case when they are not equal.

\subsubsection*{Effect of the Number of categories }

\subsubsection*{  $k_s=k$ (the number of categories is the same for all the attributes)}

We construct categories from the continuous data by setting cut points. In the first simulation setting, we assume that all the categories within each attribute have the same probability.
For example, for an attribute having four categories $k_s=4$, 
the quartiles $Q_1, Q_2, Q_3$ are used as cut points, where $Q_1, Q_2, Q_3$ are the usual lower quartile, median and upper quartile respectively, which divide the data into an equal four parts. So in this case,  $\pi_1 = \pi_2 = \pi_3 = \pi_4 = 0.25$.
In general, to create $c$ categories of a variable one needs $c-1$ cut points.

In our second simulation setting, the number of categories ($k_s$) of all the attributes is the same but the categories within each attribute may have the unequal probabilities ($\pi_c \neq 1/k$). The purpose is to investigate whether $\pi_c$ do have any effect on the imputation results.

\begin{figure}	[h]			
	\centering
	\begin{tabular}{ccc}
		& $\pi_c = 1/k$ &\\
		\includegraphics[width = 4cm, height=4cm]{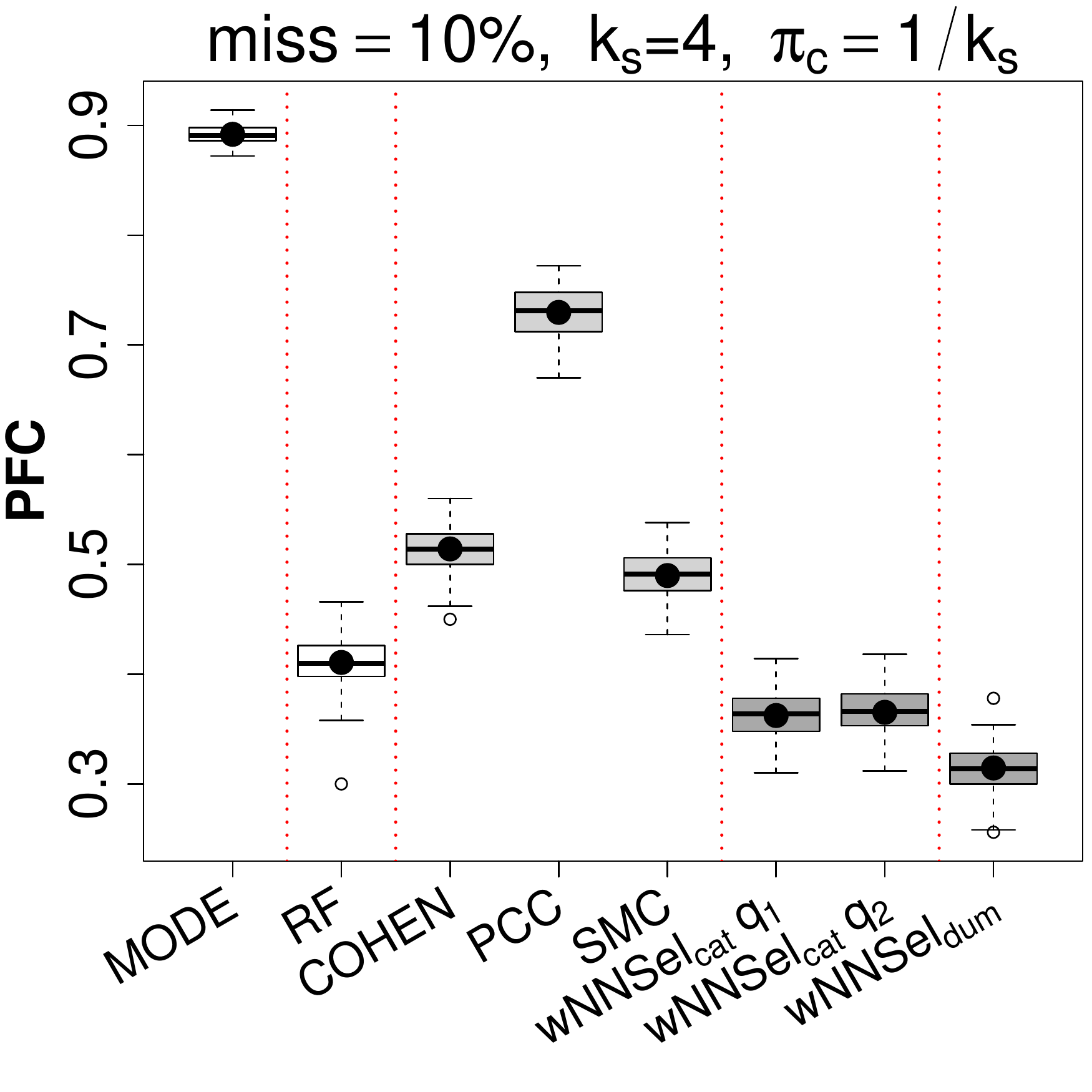} &
		\includegraphics[width = 4cm, height=4cm]{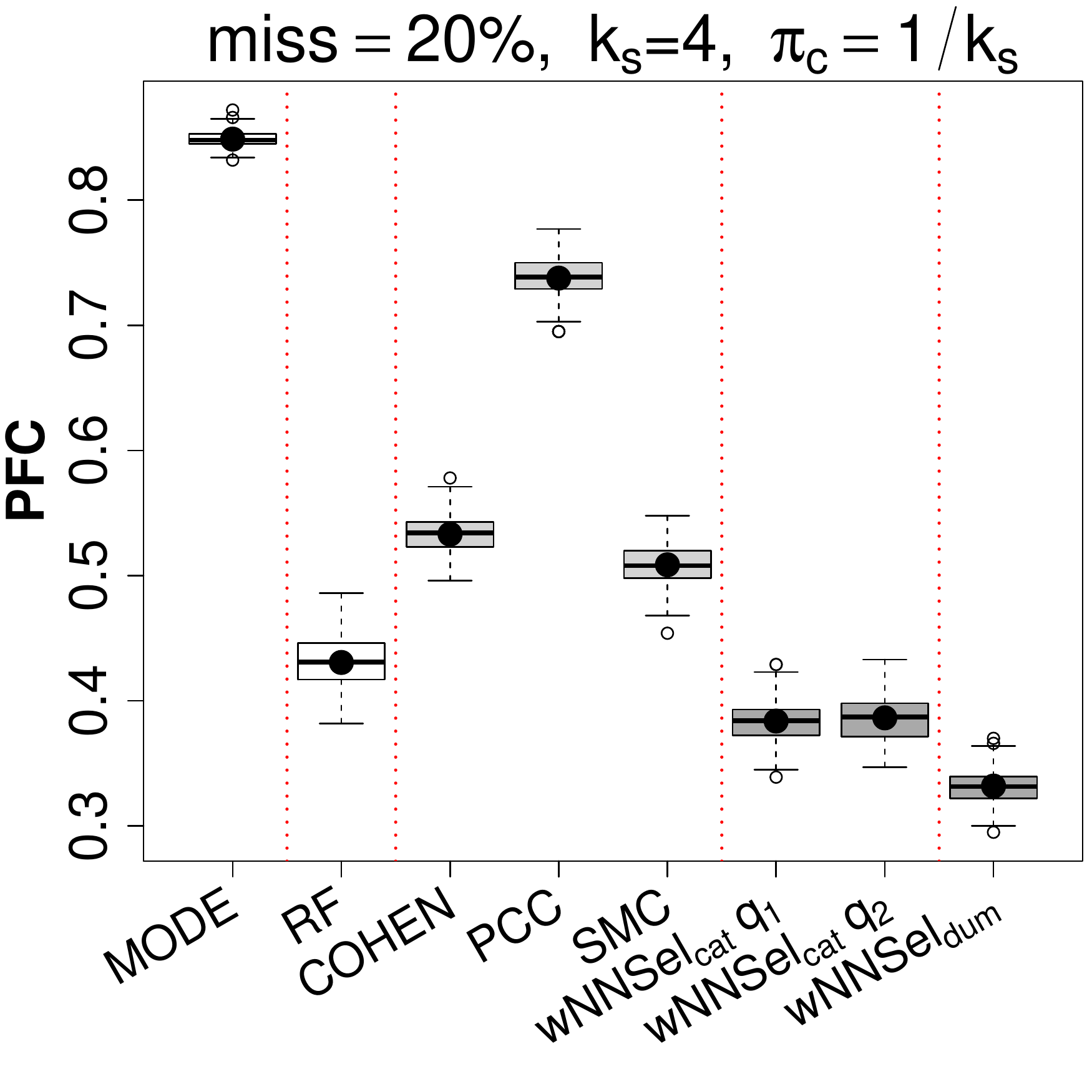}&
		\includegraphics[width = 4cm, height=4cm]{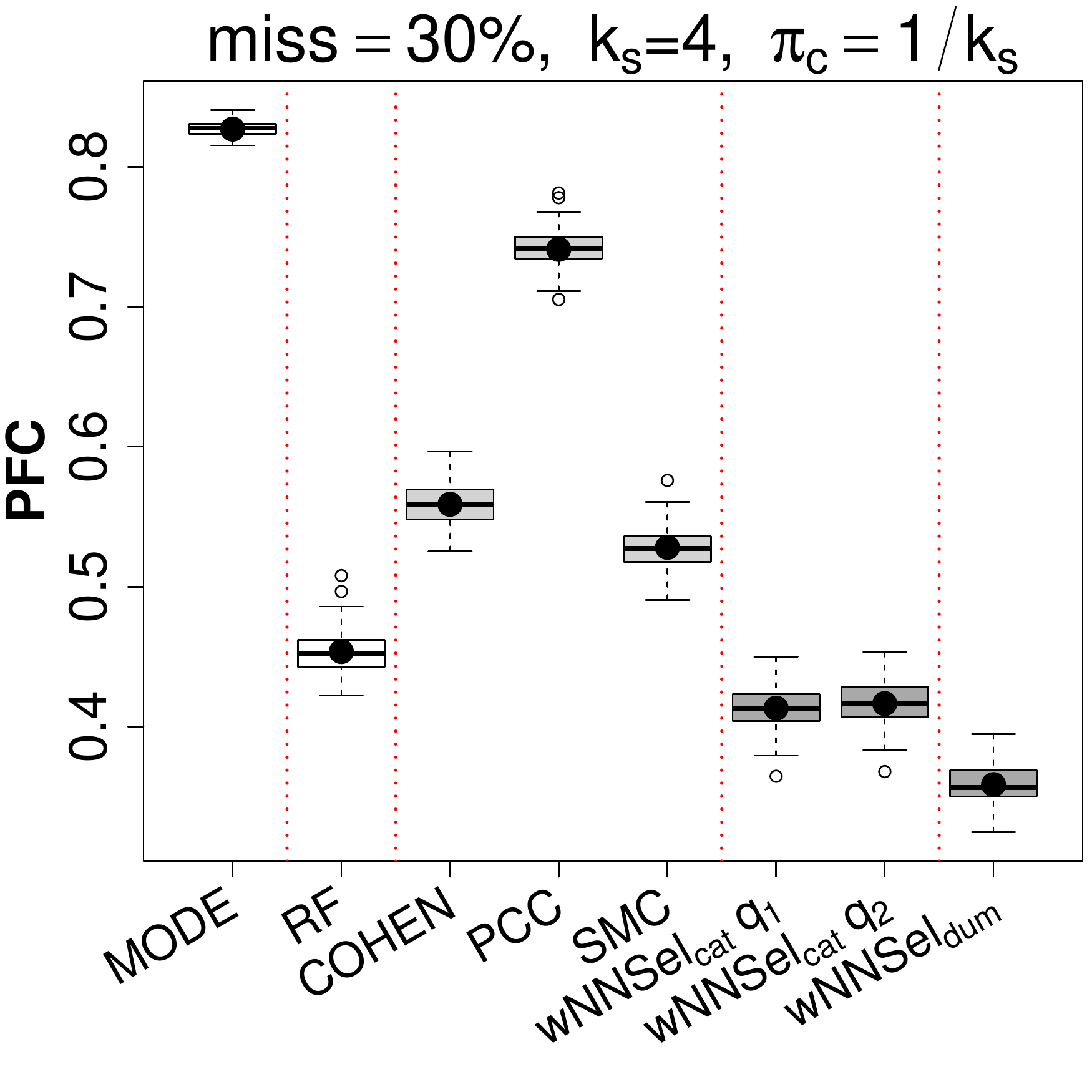} \\
		& & \\
		& $\pi_c \neq 1/k$ &\\
		\includegraphics[width = 4cm, height=4cm]{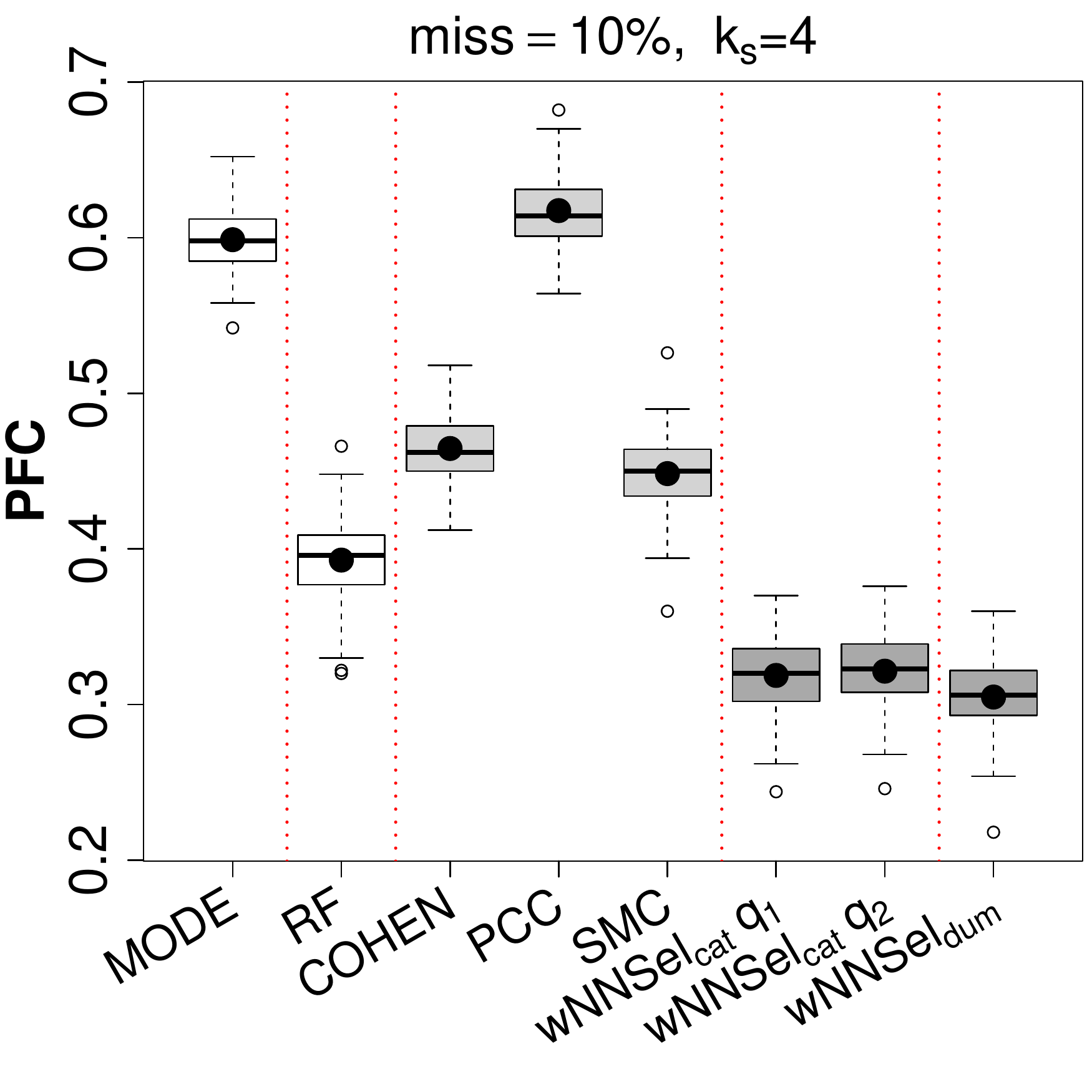}&
		\includegraphics[width = 4cm, height=4cm]{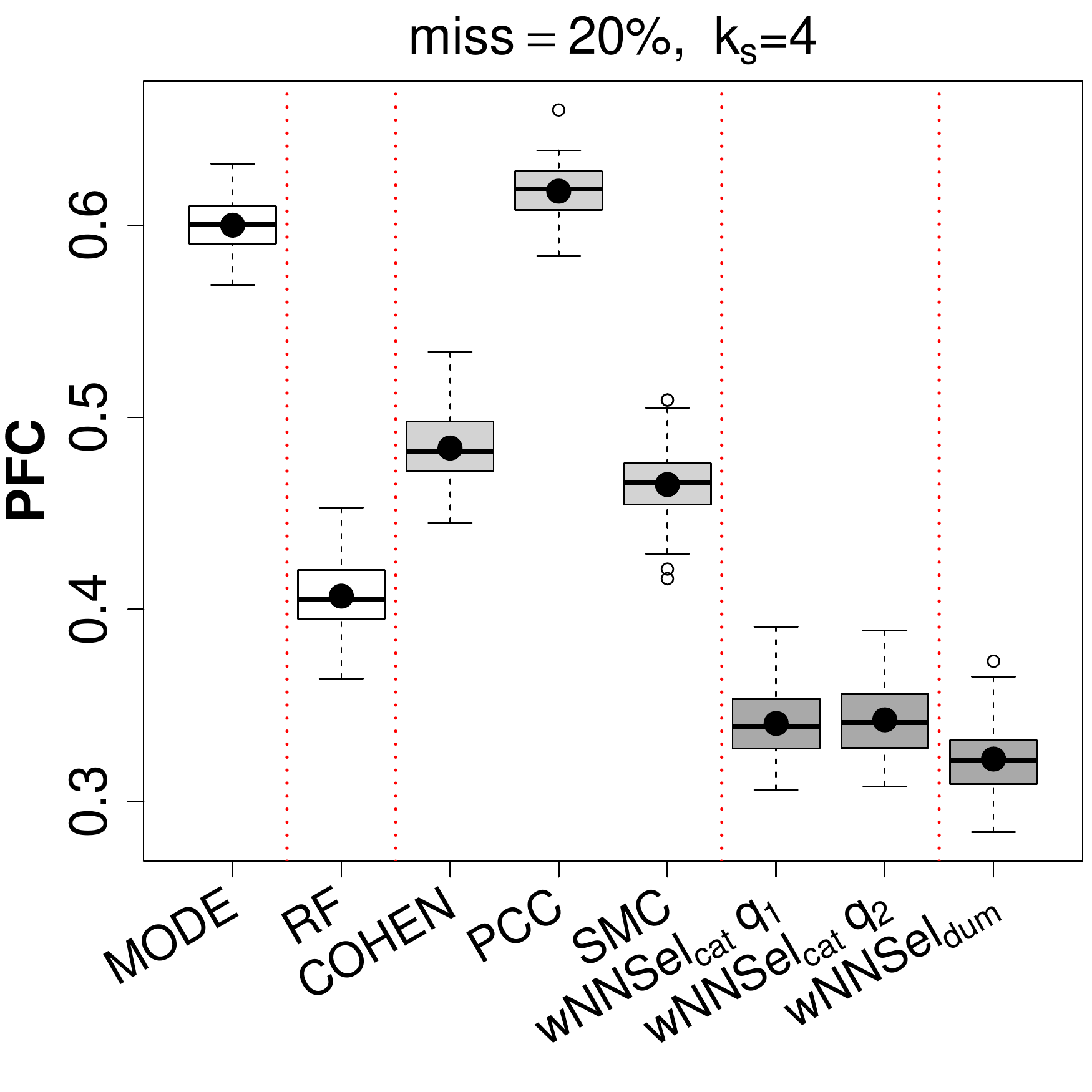}&
		\includegraphics[width = 4cm, height=4cm]{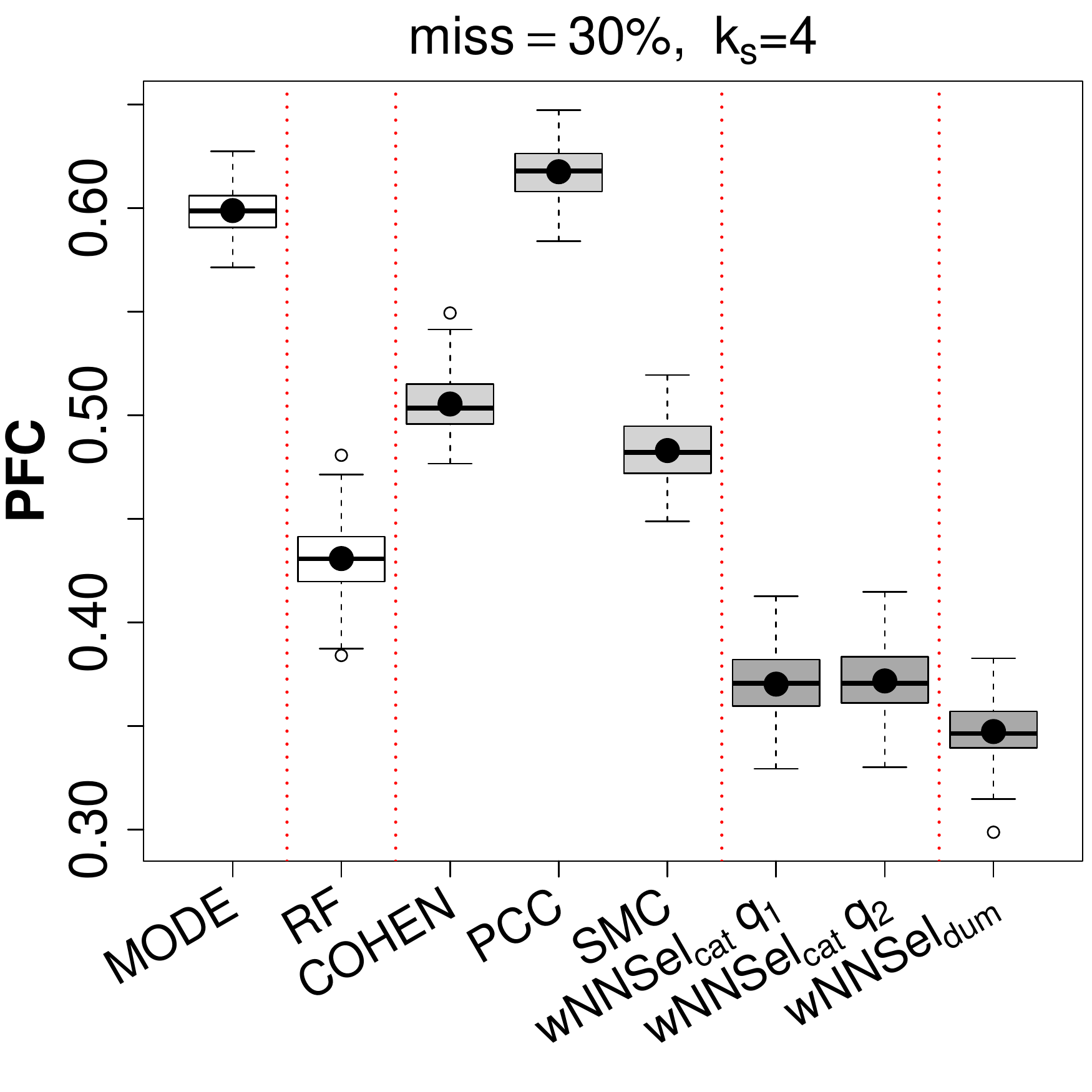} \\
		
	\end{tabular}
	\caption{Simulation study for multi-categorical data (the number of categories is the same for all the attributes): Boxplots of PFCs for MCAR missing pattern with $k_s=4$,  $S=200$ samples of size $n=100, p=50$ were drawn from a multivariate normal distribution using autoregressive  correlation structures to form the categories. Solid circles within boxes show mean values. Upper row shows when the probability of occurrence of each category is the same and lower row for probability of occurrence of each category is not same.}
	\label{fig:4Catimput}
\end{figure}

We use $q=1,2$ in the distance calculation of $\mathtt{wNNSel_{cat}}$ method to get $L_1$ and $L_2$ metrics (shown as $\mathtt{wNNSel_{cat}}\mathrm{q_1}$ and  $\mathtt{wNNSel_{cat}}\mathrm{q_2}$ in Figure \ref{fig:4Catimput} ). The tuning parameters are chosen by cross validation and these optimal values, $\lambda_{opt}$ and $m_{opt}$, are used to estimate the final imputed values. Using dummy variable method, the missing values are imputed and shown as $\mathtt{wNNSel_{dum}}$ in Figure \ref{fig:4Catimput}.  

It is seen from Figure \ref{fig:4Catimput} (upper panel), that for $\pi_c = 1/k$, $\mathtt{wNNSel_{dum}}$ method yields the smallest imputation errors followed by $\mathtt{wNNSel_{cat}}$. For the $\mathtt{wNNSel_{cat}}$ method, both values of $q$ produce similar results. 
The method by \cite{schwender2012imputing} is  also used as benchmark as all the attribute have an equal number of  categories. We used Cohen, PCC and SMC distances to compute the nearest neighbors for this method (light-gray boxes in Figure \ref{fig:4Catimput}). The PCC distance gives higher imputation errors than the Cohen and SMC distances which yield almost similar results. 
The same findings can be seen for  $\pi_c \neq 1/k$, in the lower panel of Figure \ref{fig:4Catimput}.
Overall, the replacement of missing values by the mode yields the highest errors followed by KNN and random forests, while weighted nearest neighbors imputation with weighting as proposed here provides the smallest errors.

\subsubsection*{ $k_s \neq k$  (the number of categories is different for the attributes)}

In this simulation setting, we explore whether the $\mathtt{wNNSel_{cat}}$ method works in the situation when the attributes have an unequal number of categories ($k_s \neq k$). Following the same process as in the previous subsections, we use $k_s=\{3,4\}$ and $k_s=\{3,4,5\}$ for the predictors in this case. Furthermore, the probability of occurrence of each category $(\pi_c)$ is not same i.e., $\pi_c \neq 1/k_s$. We set $n=100$, $p=50$ and $miss=10\%, 20\%, 30\%$ values are deleted at random.  The rest of the procedure of imputing missing values is the same.

\begin{figure}	[h]			
	\centering
	\begin{tabular}{cc}
		\includegraphics[scale=0.3]{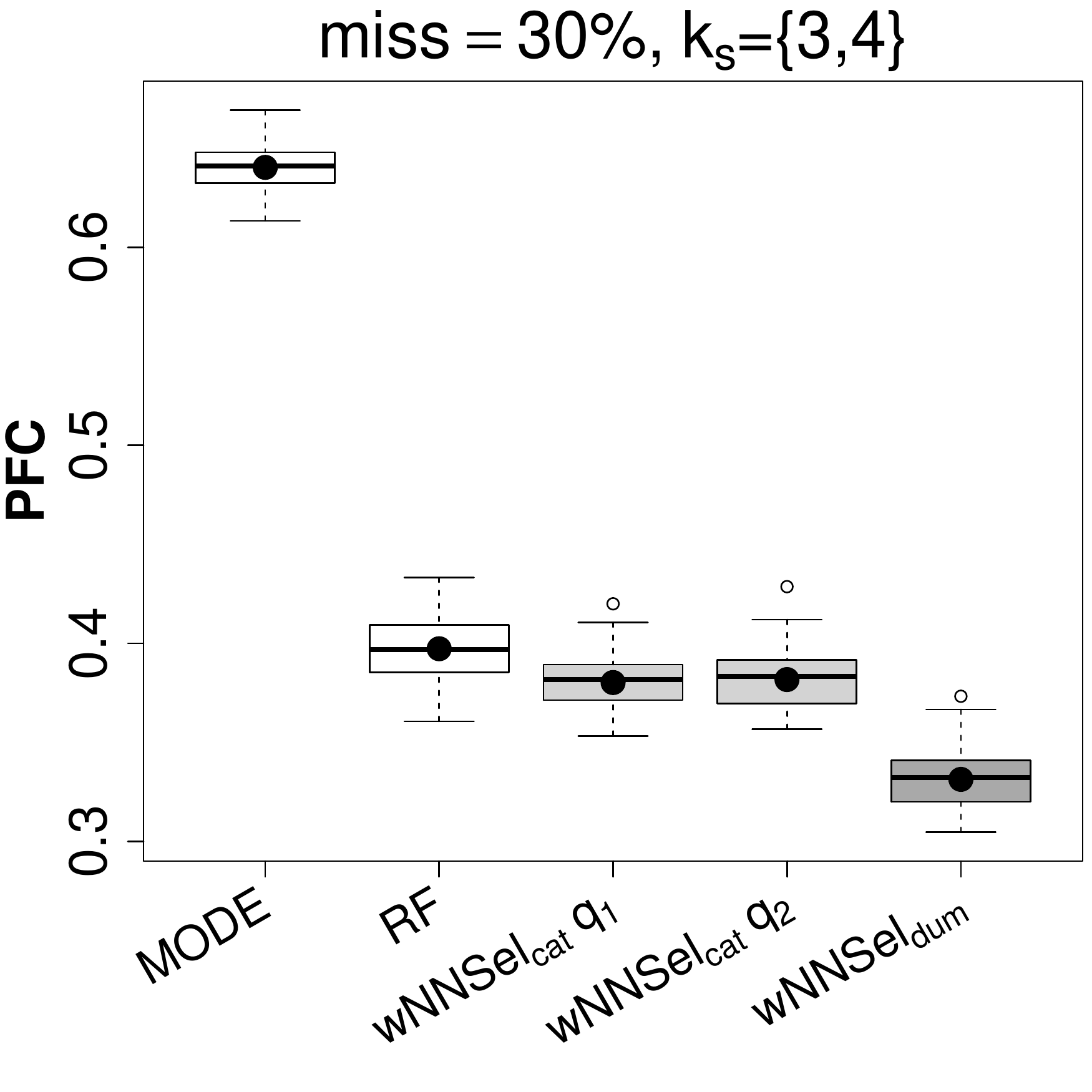}&
		\includegraphics[scale=0.3]{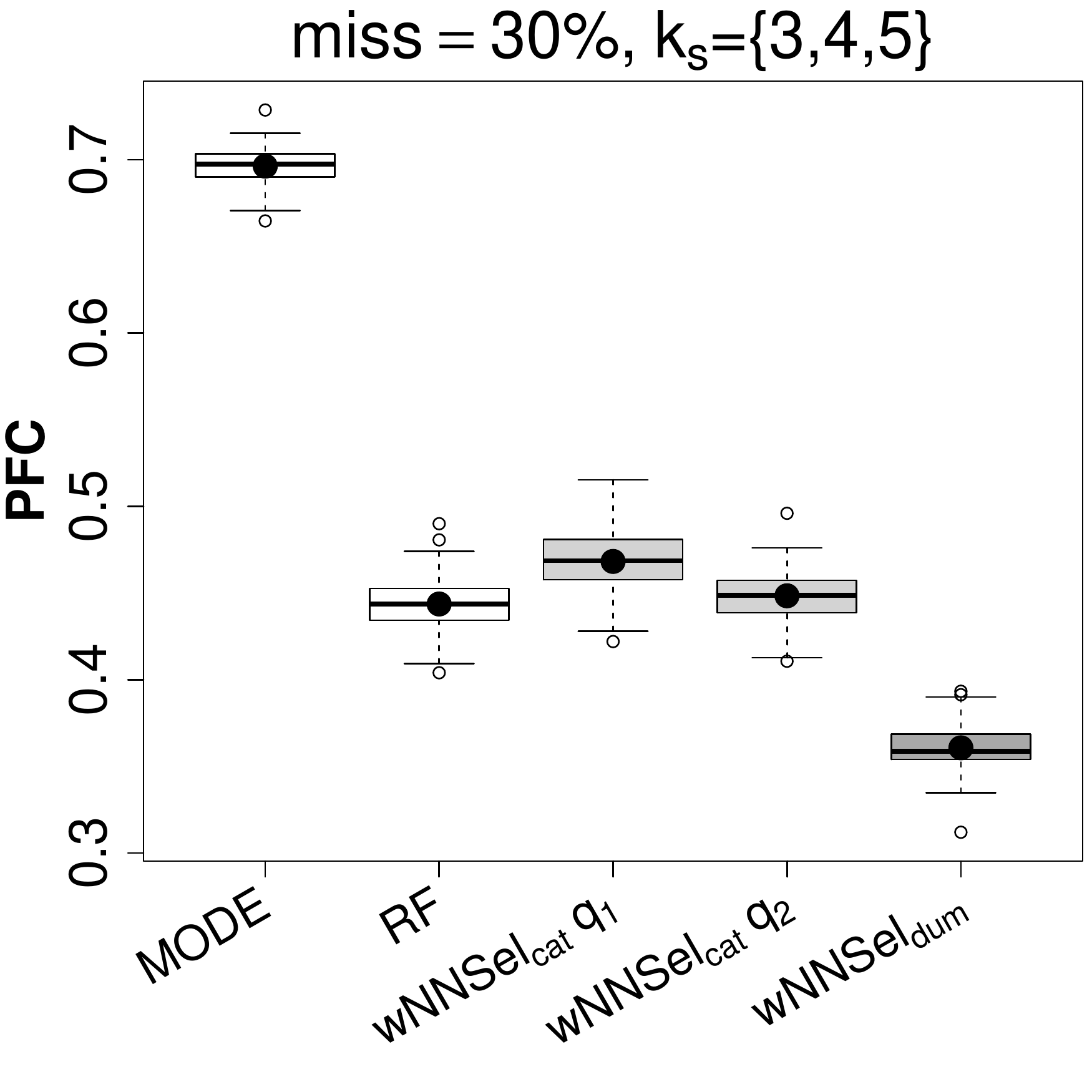} \\
	\end{tabular}
	\caption{Simulation study for multi-categorical data (the number of categories is different for the attributes): Boxplots of PFCs for MCAR missing pattern with $k_s \neq k$,  $S=200$ samples were drawn from multivariate normal distribution using autoregressive  correlation structures to form the categories.  Solid circles within boxes show mean values.}
	\label{fig:3455Catimputneq}
\end{figure}

The resulting PFCs for $miss=30\%$ only are shown in Figure \ref{fig:3455Catimputneq}. The left panel shows the results for $k_s=\{3,4\}$ and the right panel for $k_s=\{3,4,5\}$.  It is to be noted that the KNN method of \cite{schwender2012imputing} is not applicable in these settings. Clearly, the mode imputation shows the highest errors followed by random forests. It is interesting that the random forest method perform pretty well and yields similar results as the $\mathtt{wNNSel_{cat}}$ method. Here again, the smallest errors are obtained by the $\mathtt{wNNSel_{dum}}$ method in both settings considered. 
The detailed results for other the settings are shown in Table \ref{tab:catsim}.

\begin{table}[ht]
\small\sf\centering
	\caption{Comparison of imputation methods using multi-categorical simulated data}	
		\begin{tabular}{lccclrlllc}
			\toprule
			& \multirow{2}{*}{miss} 	& \multirow{2}{*}{\texttt{MODE}}		& \multirow{2}{*}{\texttt{RF}} 	& & \multicolumn{2}{c}{$\mathtt{wNNSel_{cat}}$}   & & \multirow{2}{*}{$\mathtt{wNNSel_{dum}}$} \\
			\cmidrule{6-7}
			&  		&  &  & & $q=1$&$q=2$ & & \\
			\midrule
			\multirow{3}{*}{$k_s=\{3,4\}$ } & 10\% & 0.6377 & 0.3578 & & {0.3290} & 0.3291 & & \textbf{0.3198} \\
			& 20\% & 0.6385 & 0.3767 & & {0.3524} & 0.3533 & & \textbf{0.3087} \\
			& 30 \%& 0.6404 & 0.3973 & & {0.3803} & 0.3817 & & \textbf{0.3314} \\
			& & & & & & & & \\
			
			\multirow{3}{*}{$k_s=\{3,4,5\}$} & 10\% & 0.6989 & 0.4066 & & 0.4227 & 0.3969 & & \textbf{0.3198}\\
			& 20\% & 0.6974 & 0.4212 & & 0.4399 & 0.4205 & & \textbf{0.3392}\\
			& 30 \%& 0.6963 & 0.4437 & & 0.4682 & 0.4484 & & \textbf{0.3607}\\
			\bottomrule& & 
		\end{tabular} \label{tab:catsim}
\end{table}

\subsection{Mixed (Binary and Multi-categorical) Variables}
As shown in the previous subsections that weighted imputation yields better estimates of the missing values. Specifically, $\mathtt{wNNSel_{dum}}$ performs better than $\mathtt{wNNSel_{cat}}$ in the case of the multi-categorical data, while for binary data both methods perform very similar.  
In this section we examine the performance of these methods when the data contains a mixture of binary and multi-categorical variables

\begin{figure}
	\begin{tabular}{ccc}
		\includegraphics[scale=0.25]{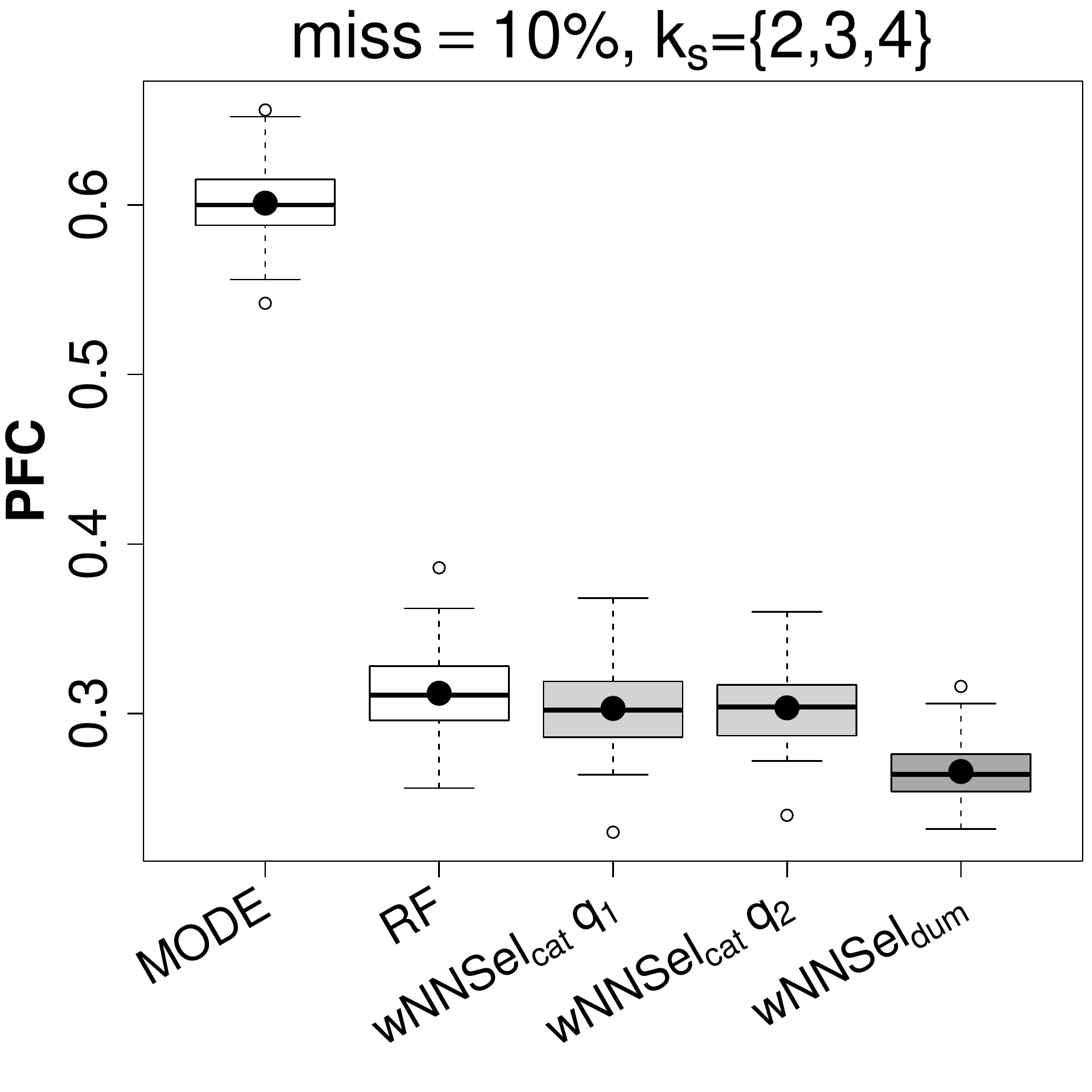}&
		\includegraphics[scale=0.25]{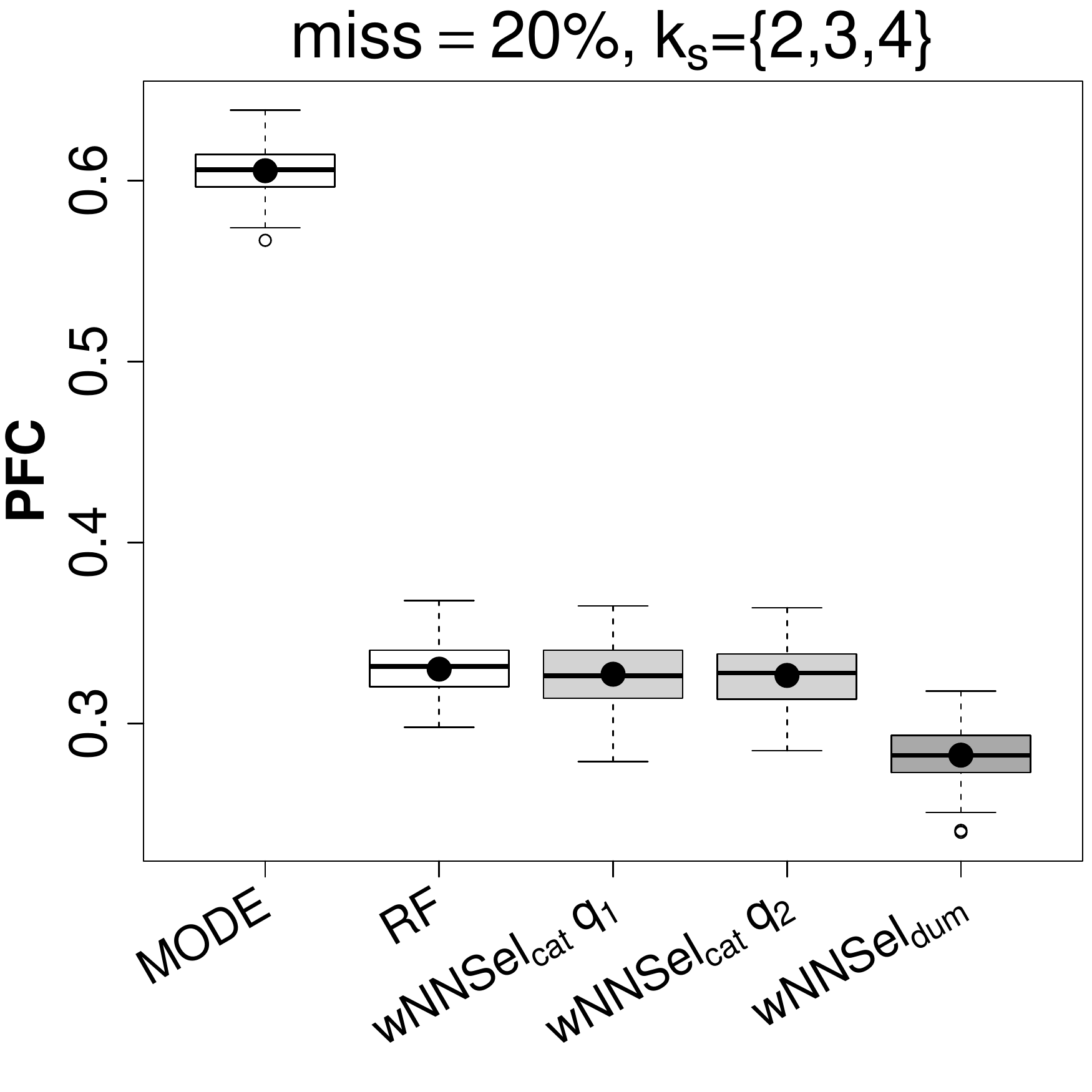}&
		\includegraphics[scale=0.25]{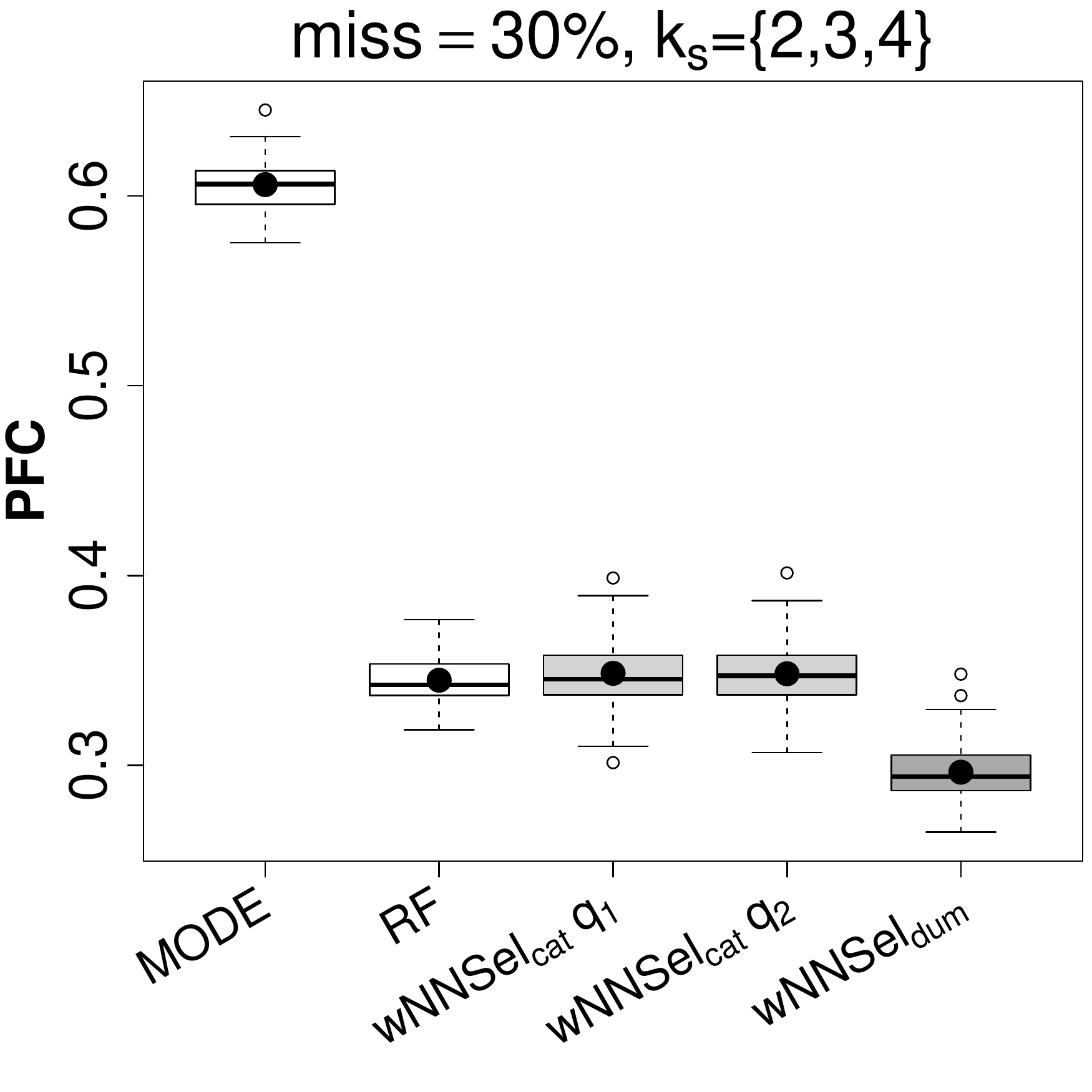}\\
	\end{tabular}
	\caption{Simulation study for mixed data: Boxplots of PFCs for MCAR missing pattern with binary and multi-categories in the data,  $S=200$ samples were drawn from multivariate normal distribution using autoregressive  correlation structures to form the categories. Solid circles within boxes show mean values.}
	\label{fig:234}	
\end{figure}

We use $k_s=\{2,3,4\}$ for $S=200$ samples of size $n=100,~ p=50$ drawn from a multivariate normal distribution using autoregressive  correlation structure. One third of the variables selected at random are converted to binary and the rest to $k_s=3,4$ categories. Then miss=10\%, 20\% and 30\% of the total values are randomly deleted to create missing values.
The rest procedure is the same as in previous subsections. The boxplots of resulting PFCs are shown in Figure \ref{fig:234}. For mixed data, the smallest imputation errors are obtained by the $\mathtt{wNNSel_{dum}}$ procedure. 
It is interesting to see that the random forest method performs as well as the $\mathtt{wNNSel_{cat}}$ method. 

The detailed results, using triangular kernel function  also, are given in Table \ref{tab:234sim}.
It is obvious that estimates using the mode yields the worst results as in the previous simulations. The random forest method provides imputation estimates that are closer to $\mathtt{wNNSel_{cat}}$. In a comparison of $\mathtt{wNNSel_{cat}}$  and random forest methods, $\mathtt{wNNSel_{cat}}$  shows slightly better results, except for 30\% missing values where the smallest average PFC=0.3484 is obtained by random forest. 
Overall, the $\mathtt{wNNSel_{dum}}$ gives the smallest PFCs in all the simulation settings considered here.

\begin{table}
	\centering \sf
	\caption{Comparison of imputation methods using binary and multi-categorical simulated data}	
	\begin{footnotesize}
		\begin{tabular} {ccccccccc}
			\toprule
			\multirow{2}{*}{miss}& \multirow{2}{*}{\texttt{MODE}}& \multirow{2}{*}{\texttt{RF}} &  &    \multicolumn{4}{c}{$\mathtt{wNNSel_{cat}}$} &  \multirow{2}{*}{$\mathtt{wNNSel_{dum}}$}\\
			\cmidrule{5-8}
			& &  &  & \texttt{Gauss.q1}  & \texttt{Gauss.q2}  & \texttt{Tri.q1} & \texttt{Tri.q2} & \\
			\midrule
			10\% & 	0.6011 &  0.3120 & & 0.3030 & 0.3034 & 0.3270 & 0.3219 & \textbf{0.2658}\\
			20\% & 	0.6054 &  0.3301 & & 0.3273 &  0.3266 & 0.3518 & 0.3464  & \textbf{0.2826}\\
			30\% & 	0.6060 &  0.3448 & & 0.3484 &  0.3482 & 0.3701 &  0.3648  & \textbf{0.2963}\\
			\bottomrule& 
		\end{tabular}
	\end{footnotesize}
	\label{tab:234sim}
\end{table}

\section{Applications} \label{sec:appli}
The results of simulation studies show that the suggested weighted nearest neighbors imputation methods ($\mathtt{wNNSel_{cat}}$ and $\mathtt{wNNSel_{dum}}$) perform better than other competitors. In this section we apply the imputation methods to real data sets. We use three different data sets with binary, multi-categorical and mixed variables.

\subsubsection*{SPECT heart data (Binary only)}

The dataset describes Single Proton Emission Computed Tomography (SPECT) images. Each of the 267 patients is classified into two categories: normal and abnormal based on $p=22$ binary feature patterns. \cite{kurgan2001knowledge} discuss this processed data set summarizing about 3000 2D SPECT images.

\subsubsection*{DNA Promoter gene sequence (Multi-categorical) }

The data for promoter instances was used by \cite{harley1987analysis} and for non-promoters by \cite{towell1990refinement}.
The total data set contains sequences of $p=57$ base pairs from $n=106$ candidates/samples. Each of the 57 variables can be grouped into one of the four DNA nucleotides; adenine, thymine, guanine or cytosine. The response variable is promoter or non-promoter instances.

\subsubsection*{Lymphography data (Binary and Multi-categorical)}

The data were obtained from n=148 patients suffering from cancer in the lymphatic of the immune system. For each patient, $p=18$ different properties were recorded on a nominal scale. Nine variables out of 18 are binary and the rest have more than two classes. Based on this information, the patients were classified into one of the four categories; normal, metastases, malign lymph or fibrosis.

\begin{figure}			
	\centering
	\begin{tabular}{ccc}
		
		\includegraphics[scale=0.25]{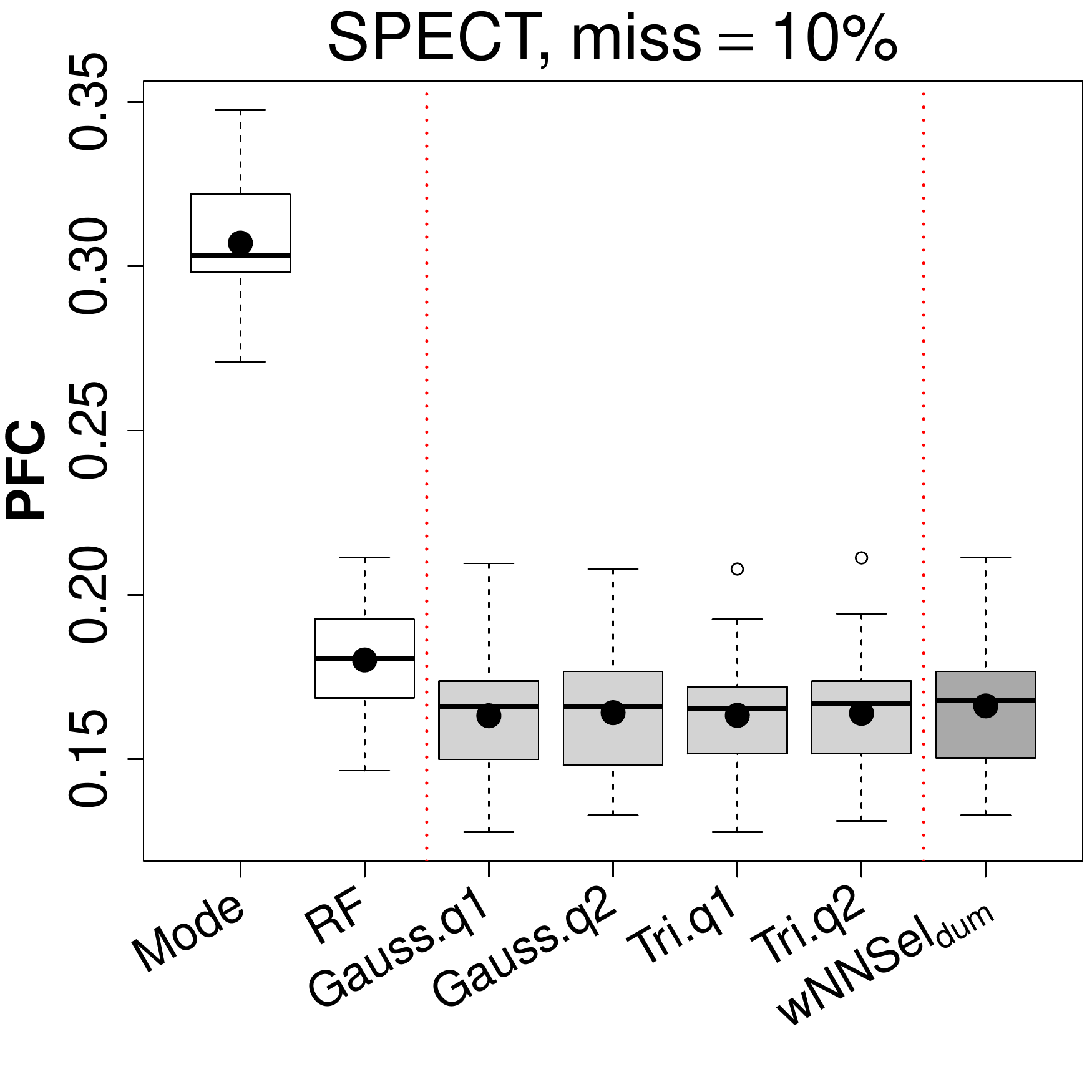} &
		\includegraphics[scale=0.25]{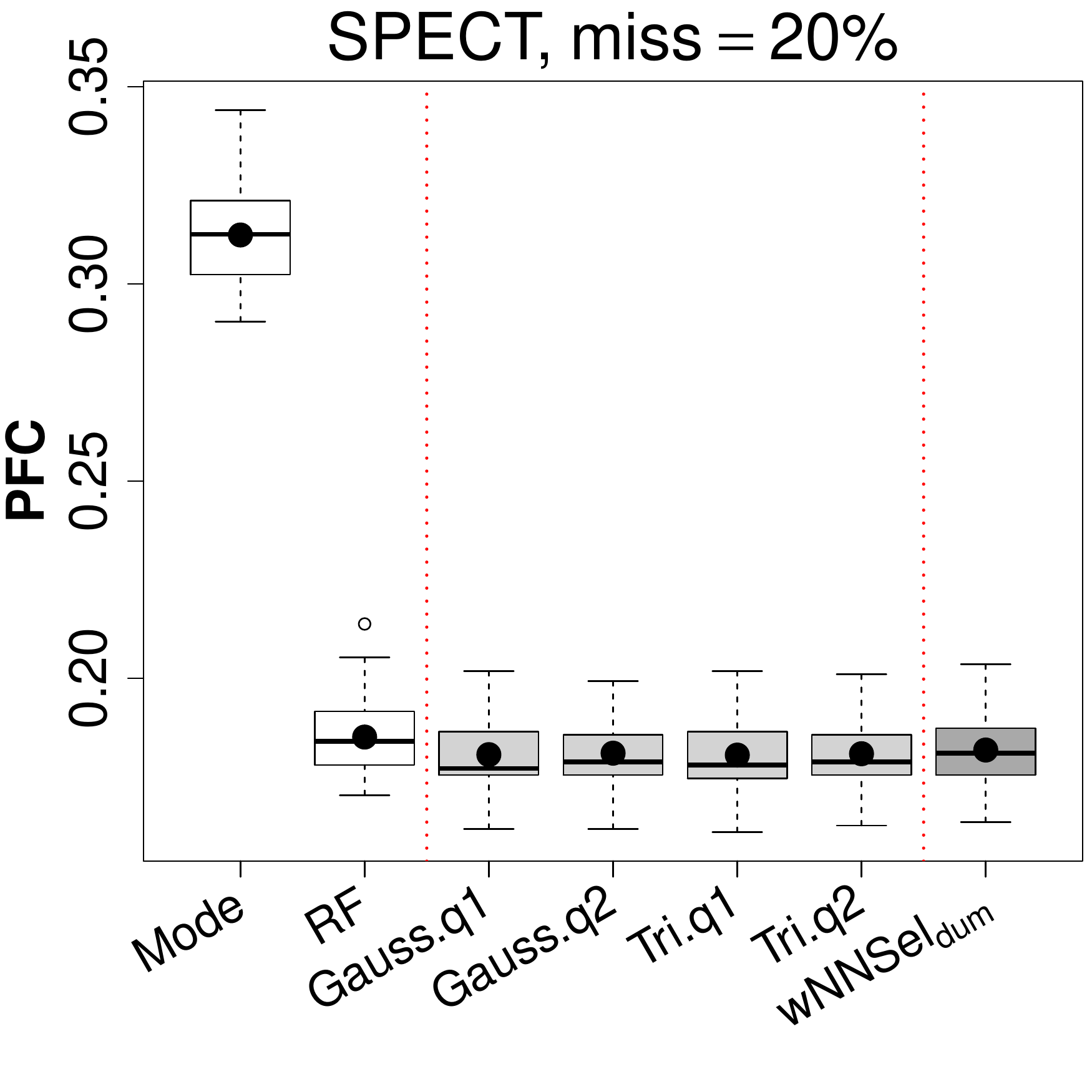} &
		\includegraphics[scale=0.25]{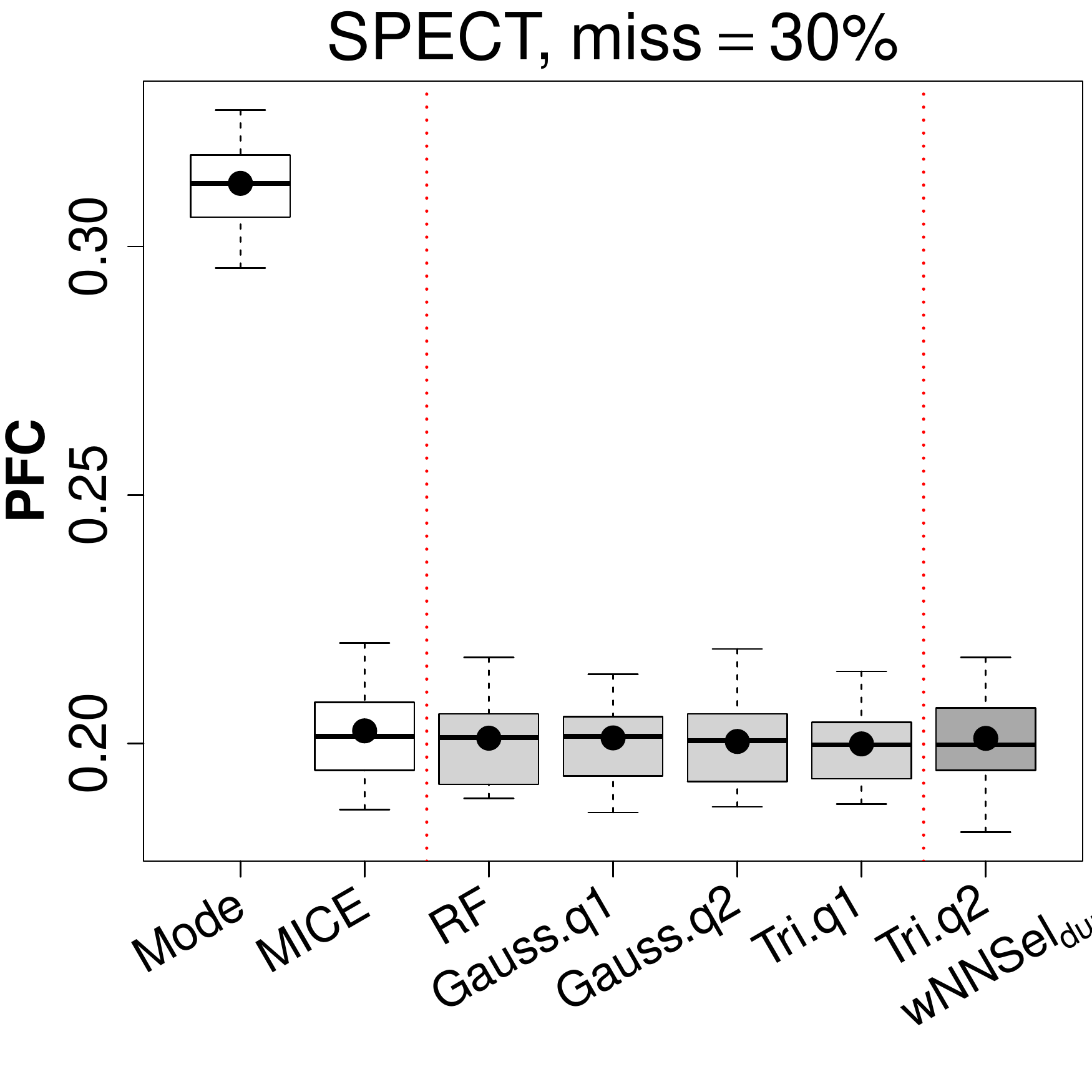} \\
		
		\includegraphics[scale=0.25]{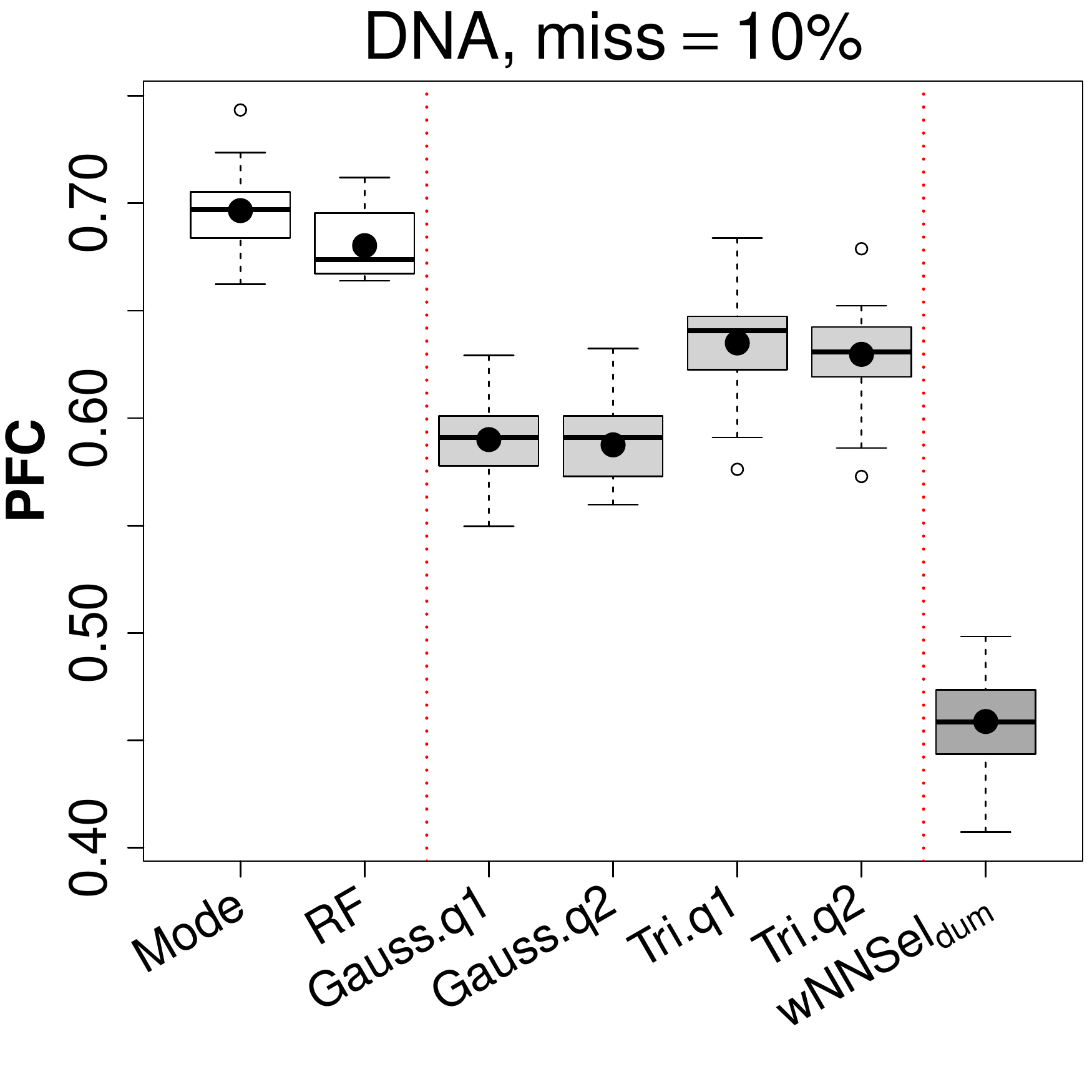} &
		\includegraphics[scale=0.25]{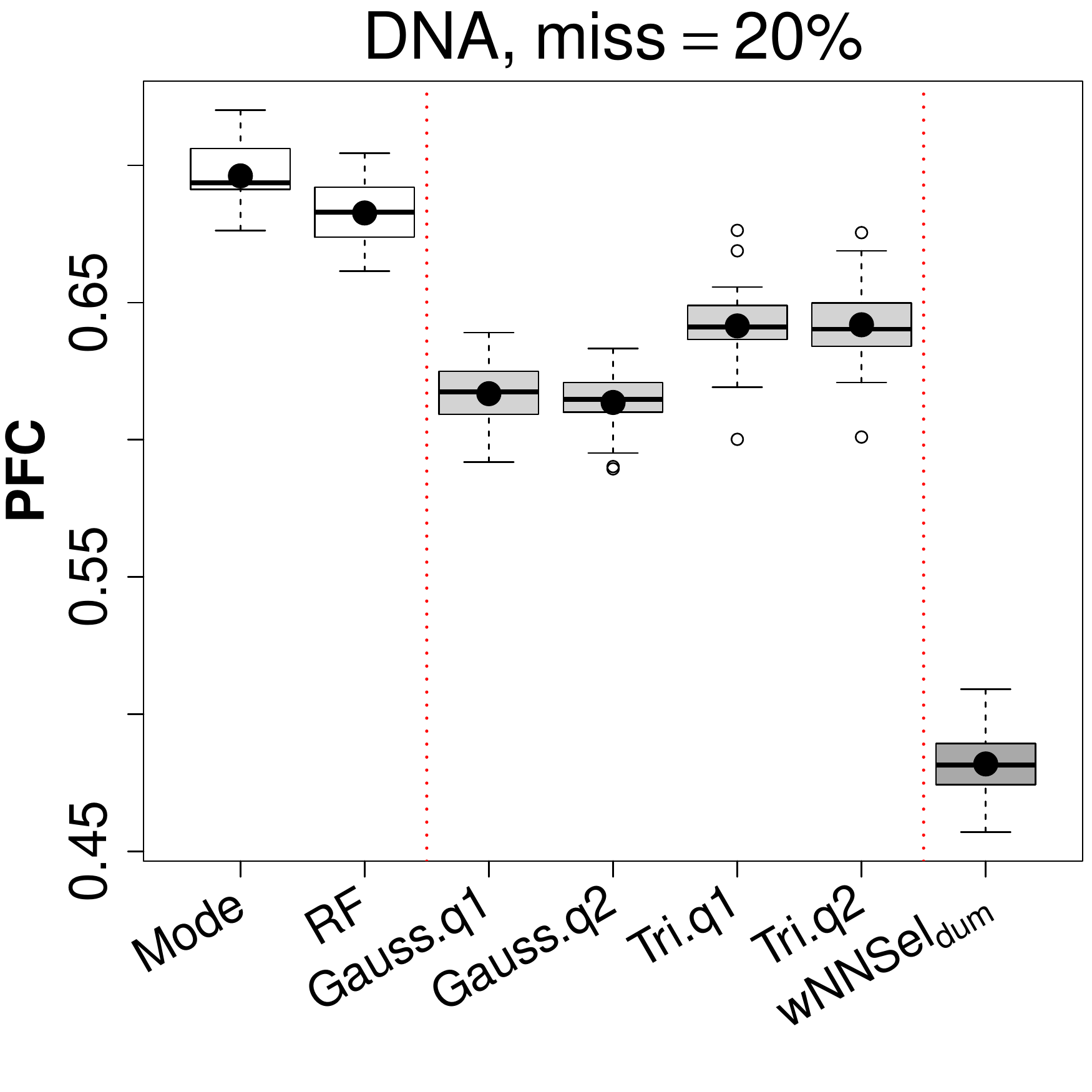} &
		\includegraphics[scale=0.25]{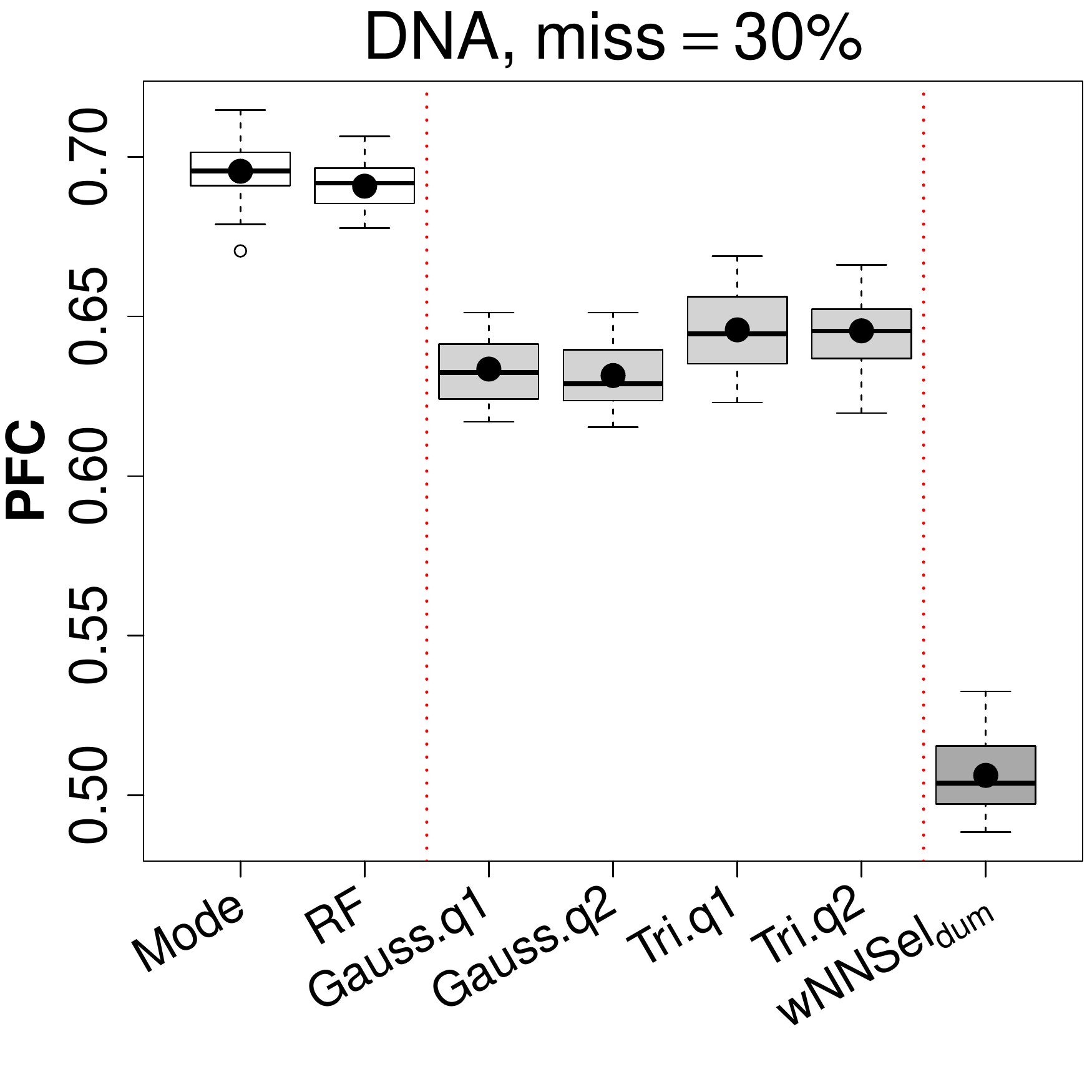} \\
		
		\includegraphics[scale=0.25]{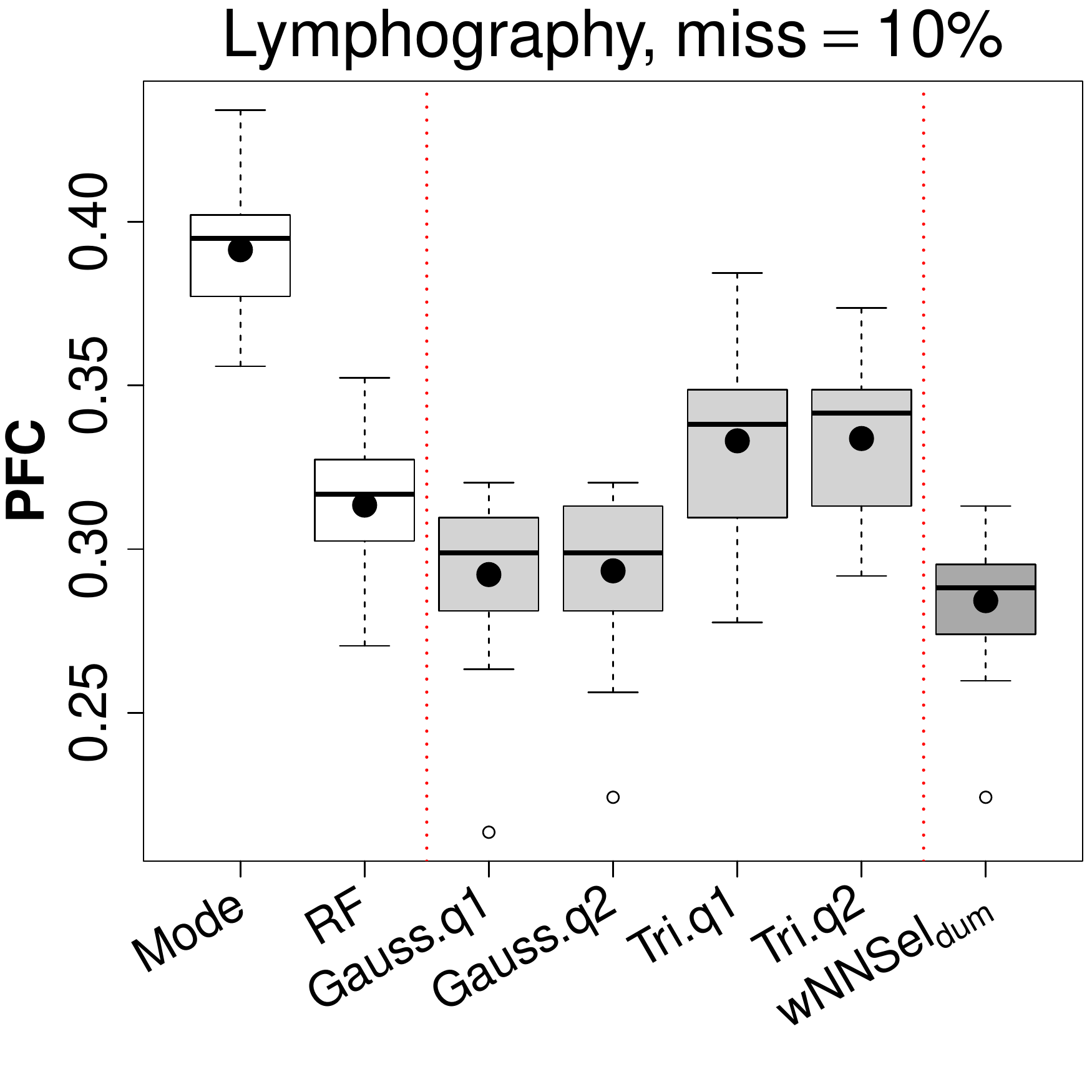} &
		\includegraphics[scale=0.25]{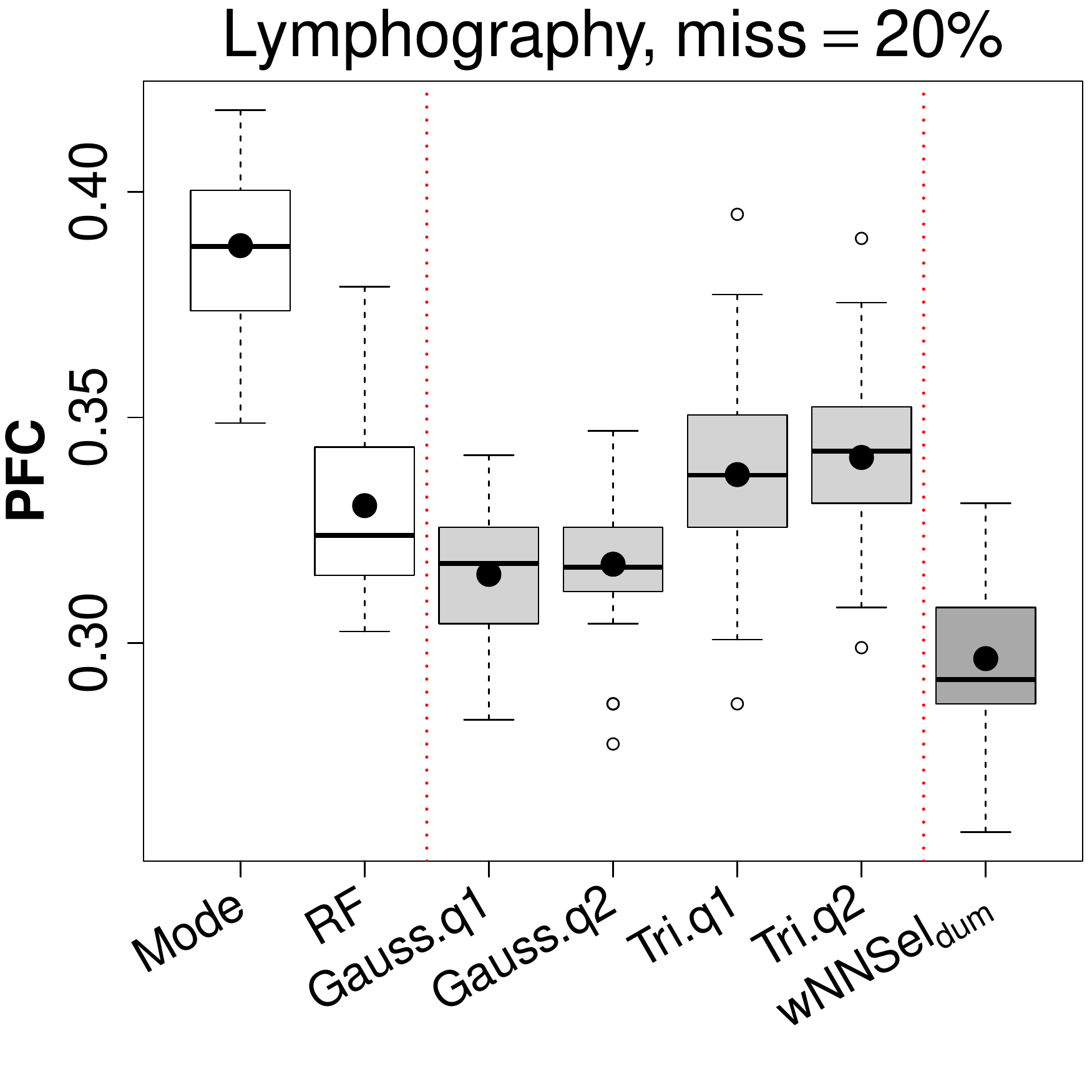} &
		\includegraphics[scale=0.25]{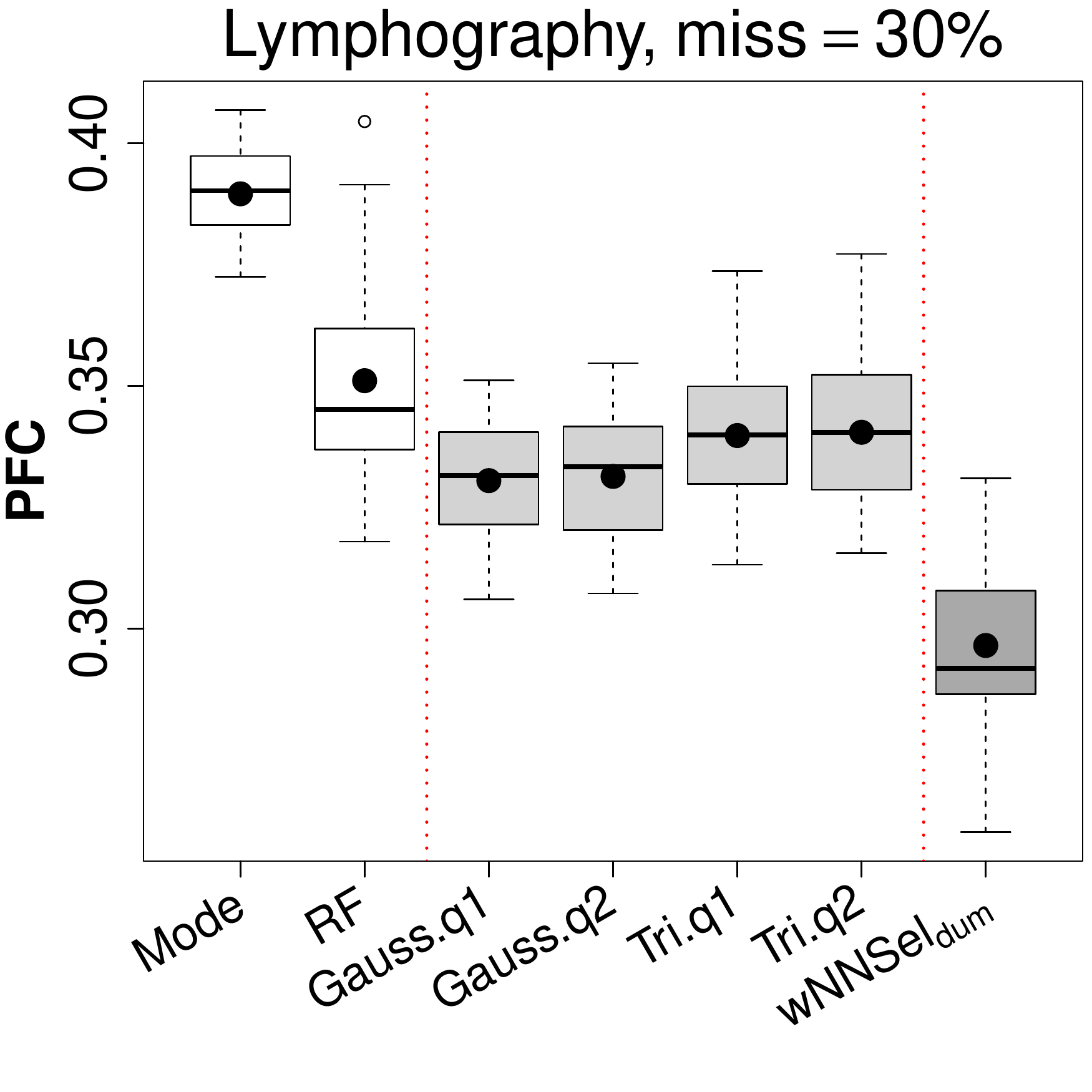} \\
		
	\end{tabular}
	\caption{Real data: Boxplots of PFCs obtained by different imputation methods. The SPECT data (upper row), DNA promoter gene sequence data (middle row) and Lymphography data (lower row) with 10\%, 20\% and 30\% missing values is shown. Grey boxes show the proposed $\mathtt{wNNSel_{cat}}$ and dark grey show  $\mathtt{wNNSel_{dum}}$ method. Solid circles within boxes show mean values. }
	\label{fig:realdata}
\end{figure}

In each data set, $miss ={10\%, 20\%, 30\%}$   values are  randomly deleted and imputation is carried out using mode, random forest,  $\mathtt{wNNSel_{cat}}$ and $\mathtt{wNNSel_{dum}}$ methods. The imputation error is computed in terms of PFC. The results of 30 independent runs are shown in Figure \ref{fig:realdata}.

For the $\mathtt{wNNSel_{cat}}$ method, we use the Gaussian and triangular kernel function each for the value $q=1,2$ (shown as \texttt{Gauss.q1}, \texttt{Gauss.q2}, \texttt{Tri.q1}, and \texttt{Tri.q2} in Figure \ref{fig:realdata}) as we intended to explore the behavior of the kernel function and the value of $q$ on the real data sets also. 
It is seen from the figure that the Gaussian kernel yields smaller PFCs as compared to PFCs obtained by using the triangular kernel for DNA and Lymphography data, while both kernels produce similar results for SPECT data. The value of $q$ does not affect the results  and produces similar PFCs. These findings confirm the simulation results obtained in the previous section. 

The strongest difference between the performance of $\mathtt{wNNSel_{cat}}$ and $\mathtt{wNNSel_{dum}}$ is seen for the DNA promoter data. 
For the SPECT heart data which contains only binary variables, neither the kernel function nor the value of $q$ have significant impact on the values of PFCs. The PFCs obtained by $\mathtt{wNNSel_{cat}}$ and $\mathtt{wNNSel_{dum}}$ are also nearly similar. These results are consistent with the previous findings from simulation studies on binary data.

The random forest method also performs well for the Lymphography data and produces  PFCs smaller than some of the $\mathtt{wNNSel_{cat}}$ methods (\texttt{Tri.q1}, and \texttt{Tri.q2}), although the smallest PFCs are obtained by $\mathtt{wNNSel_{dum}}$. 
Overall, $\mathtt{wNNSel_{dum}}$ perform better than $\mathtt{wNNSel_{cat}}$ method  for multi-categorical data, whereas both methods perform equally well in the case of binary data.

\begin{table}[h]
	\small\sf\centering
	\caption{Comparison of imputation methods using real data}	
	\resizebox{\linewidth}{!} {  
		\begin{tabular}{llcccllclcc}
			\toprule
			& & & &  \multicolumn{2}{c}{Gaussian} & &\multicolumn{2}{c}{Triangular}   & \\
			\cmidrule{5-6}  	\cmidrule{8-9}
			Data  & & \texttt{MODE}& \texttt{RF} & $q=1$  & $q=2$  &  & $q=1$ & $q=2$ & $\mathtt{wNNSel_{dum}}$ \\
			
			\midrule
			
			\rule[0.5ex]{0pt}{2.5ex}		
			SPECT & 10\% & 0.3070 & 0.1802 & \textbf{0.1632} & 0.1641 & & 0.1633 &  0.1639 & 0.1662\\
			& 20\% & 0.3124 & 0.1851 & 0.1807 & 0.1810 & & \textbf{0.1806} & 0.1809 & 0.1818 \\
			& 30\% & 0.3127 & 0.2026 & 0.2011 & 0.2012 & & 0.2004 & \textbf{0.1999} & 0.2011 \\
			\rule[0.5ex]{0pt}{3.5ex}	
			
			DNA & 10\% & 0.6966 &  0.6803 &  0.5900 & 0.5875 &  &  0.6350 &  0.6297 & \textbf{0.4586}\\
			& 20\% & 0.6962 & 0.6827 & 0.6168 & 0.6136 &  & 0.6415 & 0.6419 & \textbf{0.4818}\\
			& 30\% & 0.6955 & 0.6908 & 0.6335 & 0.6315 &  & 0.6458 & 0.6455  &  \textbf{0.5062}\\
			
			\rule[0.5ex]{0pt}{3.5ex}
			
			Lymphography & 10\% & 0.3915 &  0.3135 & {0.2922} & 0.2934 & & 0.3331 & 0.3338 &  \textbf{0.2813} \\
			& 20\% & 0.3881 &  0.3304 & {0.3152} & 0.3174 & & 0.3373 & 0.3411  &  \textbf{0.2965}\\
			& 30\% & 0.3896 &  0.3511 & {0.3305} & 0.3314 & & 0.3398 & 0.3405 & \textbf{0.3114}\\
			
			\bottomrule& 
		\end{tabular} }
		\label{tab:realdata}
	\end{table}

	\section{Concluding Remarks}
	
	We proposed a weighted distance metric based on kernel function to impute missing multi-categorical data. The method uses a distance function, called $d_{SelCat}$, that utilizes information from other covariates by taking information on association into account. To estimate the tuning parameters, a cross validation algorithm is suggested, which automatically selects the best possible values producing the smallest imputation errors. The procedure does not require a specified value of the number of nearest neighbors ($\mathsf{k}$) and provides as accurate results as the best existing methods. Simulation results show that $L_1$ and $L_2$ metrics yield similar results.  Moreover, the Gaussian kernel provided smaller imputation errors than the triangular kernel.

	To our surprise the simple method, which uses dummy variables and the classical correlation coefficient, showed the best performance.
	For binary data, both procedures $\mathtt{wNNSel_{cat}}$ and $\mathtt{wNNSel_{dum}}$ yield similar results, whereas, for multi-categorical data $\mathtt{wNNSel_{dum}}$ yields smaller imputation errors. 
	The $\mathtt{wNNSel_{dum}}$ method outperforms in simulations as well as in real data application all competitors.

\bibliography{literatur8}

\begin{thebibliography}{32}
\expandafter\ifx\csname natexlab\endcsname\relax\def\natexlab#1{#1}\fi
\expandafter\ifx\csname url\endcsname\relax
  \def\url#1{\texttt{#1}}\fi
\expandafter\ifx\csname urlprefix\endcsname\relax\def\urlprefix{URL }\fi

\bibitem[{Allison(2005)}]{allison2005imputation}
Allison, P.~D., 2005. Imputation of categorical variables with proc mi. SUGI 30
  proceedings 113~(30), 1--14.

\bibitem[{Andridge and Little(2010)}]{andridge2010review}
Andridge, R.~R., Little, R.~J., 2010. A review of hot deck imputation for
  survey non-response. International statistical review 78~(1), 40--64.

\bibitem[{Breiman(2001)}]{breiman2001random}
Breiman, L., 2001. Random forests. Machine learning 45~(1), 5--32.

\bibitem[{Chen and Shao(2000)}]{chen2000nearest}
Chen, J., Shao, J., 2000. Nearest neighbor imputation for survey data. Journal
  of official statistics 16~(2), 113.

\bibitem[{Cohen(1960)}]{cohen1960kappa}
Cohen, J., 1960. A coefficient of agreement for nominal scales. Educational and
  Psychosocial Measurement 20, 37--46.

\bibitem[{Cram{\'e}r(1946)}]{cramer1946methods}
Cram{\'e}r, H., 1946. Methods of mathematical statistics. Princeton: Princeton
  Univer-sity Press. CramerMethods of Mathematical Statistics1946.

\bibitem[{Cranmer and Gill(2013)}]{cranmer2013we}
Cranmer, S.~J., Gill, J., 2013. We have to be discrete about this: A
  non-parametric imputation technique for missing categorical data. British
  Journal of Political Science 43~(02), 425--449.

\bibitem[{Eisemann et~al.(2011)Eisemann, Waldmann, and
  Katalinic}]{eisemann2011imputation}
Eisemann, N., Waldmann, A., Katalinic, A., 2011. Imputation of missing values
  of tumour stage in population-based cancer registration. BMC medical research
  methodology 11~(1), 1.

\bibitem[{Erosheva et~al.(2002)Erosheva, Fienberg, and
  Junker}]{erosheva2002alternative}
Erosheva, E.~A., Fienberg, S.~E., Junker, B.~W., 2002. Alternative statistical
  models and representations for large sparse multi-dimensional contingency
  tables. In: Annales de la Facult{\'e} des sciences de Toulouse:
  Math{\'e}matiques. Vol.~11. pp. 485--505.

\bibitem[{Ezzati-Rice et~al.(1995)Ezzati-Rice, Johnson, Khare, Little, Rubin,
  and Schafer}]{ezzati1995simulation}
Ezzati-Rice, T.~M., Johnson, W., Khare, M., Little, R.~J., Rubin, D.~B.,
  Schafer, J.~L., 1995. A simulation study to evaluate the performance of
  model-based multiple imputations in nchs health examination surveys. In:
  Proceedings of the Annual research Conference. Vol. 257266.

\bibitem[{Harley and Reynolds(1987)}]{harley1987analysis}
Harley, C.~B., Reynolds, R.~P., 1987. Analysis of e. coli pormoter sequences.
  Nucleic acids research 15~(5), 2343--2361.

\bibitem[{Hill(2012)}]{hill2012four}
Hill, J., 2012. Four techniques for dealing with missing data in criminal
  justice. In: annual meeting of the ASC Annual Meeting, Palmer House Hilton,
  Chicago, IL.

\bibitem[{Horton et~al.(2003)Horton, Lipsitz, and Parzen}]{horton2003potential}
Horton, N.~J., Lipsitz, S.~R., Parzen, M., 2003. A potential for bias when
  rounding in multiple imputation. The American Statistician 57~(4), 229--232.

\bibitem[{Kurgan et~al.(2001)Kurgan, Cios, Tadeusiewicz, Ogiela, and
  Goodenday}]{kurgan2001knowledge}
Kurgan, L.~A., Cios, K.~J., Tadeusiewicz, R., Ogiela, M., Goodenday, L.~S.,
  2001. Knowledge discovery approach to automated cardiac spect diagnosis.
  Artificial intelligence in medicine 23~(2), 149--169.

\bibitem[{Liao et~al.(2014)Liao, Lin, Kang, Chandra, Bon, Kaminski, Sciurba,
  and Tseng}]{liao2014missing}
Liao, S.~G., Lin, Y., Kang, D.~D., Chandra, D., Bon, J., Kaminski, N., Sciurba,
  F.~C., Tseng, G.~C., 2014. Missing value imputation in high-dimensional
  phenomic data: imputable or not, and how? BMC bioinformatics 15~(1), 346.

\bibitem[{Little and Rubin(2014)}]{little2014statistical}
Little, R.~J., Rubin, D.~B., 2014. Statistical analysis with missing data. John
  Wiley \& Sons.

\bibitem[{Pantanowitz and Marwala(2009)}]{pantanowitz2009missing}
Pantanowitz, A., Marwala, T., 2009. Missing data imputation through the use of
  the random forest algorithm. In: Advances in Computational Intelligence.
  Springer, pp. 53--62.

\bibitem[{Rieger et~al.(2010)Rieger, Hothorn, and Strobl}]{rieger2010random}
Rieger, A., Hothorn, T., Strobl, C., 2010. Random forests with missing values
  in the covariates.

\bibitem[{Rubin(1987)}]{rubin1987multiple}
Rubin, D.~B., 1987. Multiple imputation for nonresponse in surveys. New York:
  Wiley.

\bibitem[{Rubin and Schenker(1986)}]{RubinSchenker1986}
Rubin, D.~B., Schenker, N., 1986. Multiple imputation for interval estimation
  from simple random samples with ignorable nonresponse. Journal of the
  American Statistical Association 81~(394), 366--374.

\bibitem[{Schafer(1997)}]{schafer1997analysis}
Schafer, J.~L., 1997. Analysis of incomplete multivariate data. CRC press.

\bibitem[{Schafer and Graham(2002)}]{SchaferGramham2002}
Schafer, J.~L., Graham, J.~W., 2002. Missing data: our view of the state of the
  art. Psychological methods 7~(2), 147.

\bibitem[{Schwender(2012)}]{schwender2012imputing}
Schwender, H., 2012. Imputing missing genotypes with weighted k nearest
  neighbors. Journal of Toxicology and Environmental Health, Part A 75~(8-10),
  438--446.

\bibitem[{Schwender and Fritsch(2013)}]{RPackage:scrime}
Schwender, H., Fritsch, A., 2013. scrime: Analysis of High-Dimensional
  Categorical Data such as SNP Data. R package version 1.3.3.
\newline\urlprefix\url{http://CRAN.R-project.org/package=scrime}

\bibitem[{Segal(2004)}]{segal2004machine}
Segal, M.~R., 2004. Machine learning benchmarks and random forest regression.
  Center for Bioinformatics \& Molecular Biostatistics.

\bibitem[{Sokal and Michener(1958)}]{sokal1958statistical}
Sokal, R.~R., Michener, C.~D., 1958. A statistical method for evaluating
  systematic relationships. The University of Kansas science bulletin 38,
  1409--1438.

\bibitem[{Stekhoven(2013)}]{RPackage:missForest}
Stekhoven, D.~J., 2013. missForest: Nonparametric Missing Value Imputation
  using Random Forest. R package version 1.4.

\bibitem[{Stekhoven and B{\"u}hlmann(2012)}]{stekhoven2012missforest}
Stekhoven, D.~J., B{\"u}hlmann, P., 2012. Missforest: non-parametric missing
  value imputation for mixed-type data. Bioinformatics 28~(1), 112--118.

\bibitem[{Templ et~al.(2016)Templ, Alfons, Kowarik, and
  Prantner}]{RPackage:vim}
Templ, M., Alfons, A., Kowarik, A., Prantner, B., 2016. VIM: Visualization and
  Imputation of Missing Values. R package version 4.5.0.
\newline\urlprefix\url{https://CRAN.R-project.org/package=VIM}

\bibitem[{Towell et~al.(1990)Towell, Shavlik, and
  Noordewier}]{towell1990refinement}
Towell, G.~G., Shavlik, J.~W., Noordewier, M.~O., 1990. Refinement of
  approximate domain theories by knowledge-based neural networks. In:
  Proceedings of the eighth National conference on Artificial intelligence.
  Boston, MA, pp. 861--866.

\bibitem[{Troyanskaya et~al.(2001)Troyanskaya, Cantor, Sherlock, Brown, Hastie,
  Tibshirani, Botstein, and Altman}]{troyanskaya:2001}
Troyanskaya, O., Cantor, M., Sherlock, G., Brown, P., Hastie, T., Tibshirani,
  R., Botstein, D., Altman, R.~B., 2001. Missing value estimation methods for
  dna microarrays. Bioinformatics 17~(6), 520--525.

\bibitem[{Tutz and Ramzan(2015)}]{TuRam2015}
Tutz, G., Ramzan, S., 2015. Improved methods for the imputation of missing data
  by nearest neighbor methods. Computational Statistics \& Data Analysis,
  84--99.

\end{thebibliography}

\end{document}